\documentclass[final,3p,times]{elsarticle}
\usepackage{graphicx}
\usepackage{amssymb}
\usepackage{amsmath}
\usepackage{hyperref}

\journal{Computer Physics Communications}

\vspace*{-6mm}
\begin{document}

\begin{frontmatter}

\title{C and Fortran OpenMP programs for rotating Bose-Einstein condensates}

\author[col]{Ramavarmaraja Kishor Kumar}
\ead{kishor.bec@gmail.com}

\author[scl]{Vladimir Lon\v{c}ar}
\ead{vladimir.loncar@ipb.ac.rs}

\author[bdu]{Paulsamy Muruganandam}
\ead{anand@bdu.ac.in}

\author[ift]{Sadhan K. Adhikari\corref{author}}
\ead{sk.adhikari@unesp.br}

\author[scl]{Antun Bala\v{z}}
\ead{antun.balaz@ipb.ac.rs}

\cortext[author]{Corresponding author.}
\address[col]{Instituto de F\'isica, Universidade de S\~ao Paulo, 05508-090 S\~ao Paulo, Brazil}
\address[scl]{Scientific Computing Laboratory, Center for the Study of Complex Systems, Institute of Physics Belgrade, University of Belgrade, Serbia}
\address[bdu]{Department of Physics, Bharathidasan University, Palkalaiperur Campus, Tiruchirappalli -- 620024, Tamil Nadu, India}
\address[ift]{Instituto de F\'{\i}sica Te\'{o}rica, UNESP -- Universidade Estadual Paulista, 01.140-70 S\~{a}o Paulo, S\~{a}o Paulo, Brazil}

\begin{abstract}
We present OpenMP versions of C and Fortran programs for solving the Gross-Pitaevskii equation for a rotating trapped Bose-Einstein condensate (BEC) in two (2D) and three (3D) spatial dimensions. The programs can be used to generate vortex lattices and study dynamics of rotating BECs. We use the split-step Crank-Nicolson algorithm for imaginary- and real-time propagation to calculate stationary states and BEC dynamics, respectively. The simulation input parameters for the C programs are provided via input files, while for the Fortran programs they are given at the beginning of each program and therefore their change requires recompilation of the corresponding program. The programs propagate the condensate wave function and calculate several relevant physical quantities, such as the energy, the chemical potential, and the root-mean-square sizes. {The imaginary-time propagation starts with an analytic wave function with one vortex at the trap center, modulated by a random phase at different space points.} Nevertheless, the converged wave function for a rapidly rotating BEC with a large number of vortices is most efficiently calculated using the pre-calculated converged wave function of a rotating BEC containing a smaller number of vortices as the initial state rather than using an analytic wave function with one vortex as the initial state. These pre-calculated initial states exhibit rapid convergence for fast-rotating condensates to states containing multiple vortices with an appropriate phase structure. This is illustrated here by calculating vortex lattices with up to 61 vortices in 2D and 3D. Outputs of the programs include calculated physical quantities, as well as the wave function and different density profiles (full density, integrated densities in lower dimensions, and density cross-sections). The provided real-time propagation programs can be used to study the dynamics of a rotating BEC using the imaginary-time stationary wave function as the initial state. We also study the efficiency of parallelization of the present OpenMP C and Fortran programs with different compilers.
\end{abstract}

\begin{keyword}
Rotating Bose-Einstein condensate; Gross-Pitaevskii equation; Split-step Crank-Nicolson scheme;
C programs; Fortran programs; OpenMP; Partial differential equation; Vortex lattice
\end{keyword}

\end{frontmatter}

\begin{small}
\noindent
{\bf Program summary}
\noindent\\
{\em Program title:} BEC-GP-ROT-OMP, consisting of: (1) BEC-GP-ROT-OMP-C package, containing programs (i) bec-gp-rot-2d-th and (ii) bec-gp-rot-3d-th; (2) BEC-GP-ROT-OMP-F package, containing programs (i) bec-gp-rot-2d-th and (ii) bec-gp-rot-3d-th.
\noindent\\
{\em Program Files doi:} \href{https://doi.org/10.17632/cw7tkn22v2.2}{https://doi.org/10.17632/cw7tkn22v2.2}\\
{\em Licensing provisions:} Apache License 2.0\\
{\em Programming language:} OpenMP C; OpenMP Fortran. The C programs are tested with the GNU, Intel, PGI, Oracle, and Clang compiler, and the Fortran programs are tested with the GNU, Intel, PGI, and Oracle compiler.
\noindent\\
{\em Nature of problem:}
The present Open Multi-Processing (OpenMP) C and Fortran programs solve the time-dependent nonlinear partial differential Gross-Pitaevskii (GP) equation for a trapped rotating Bose-Einstein condensate in two (2D) and three (3D) spatial dimensions in a fully anisotropic traps.\\
{\em Solution method:}
We employ the split-step Crank-Nicolson algorithm to discretize the time-dependent GP equation in space and time. The discretized equation is then solved by imaginary- or real-time propagation, employing adequately small space and time steps, to yield the solution of stationary and non-stationary problems, respectively.
\end{small}

\newpage
\section{Introduction}
\label{sec:intro}

Previously published Fortran \cite{bec2009} and C \cite{bec2012} programs, and their OpenMP extensions \cite{bec2016,bec2017x} are now popular tools for solving the Gross-Pitaevskii (GP) equation and are enjoying widespread use. These programs have been later extended to the more complex scenario of dipolar atoms \cite{dbec2015}. The C programs have been adapted to run even faster on modern multi-core computers using general-purpose graphic processing units with Nvidia CUDA and computer clusters using Message Passing Interface (MPI) \cite{dbec2016}. In this paper, we present new OpenMP C and Fortran programs to solve the GP equation 
for a rotating trapped Bose-Einstein condensate (BEC) and to generate a vortex lattice, based on our earlier work \cite{bec2016,bec2017x}. This is a problem of general interest for both theoreticians \cite{fetter,fetter2} and experimentalists \cite{expt}.

The GP equation for a rotating trapped BEC can be conveniently solved by the imaginary- \cite{feder,zeng,danaila} and real-time evolution \cite{ueda} methods. The solution algorithms rely on transforming the GP equation to the rotating frame, where the 
rotating BEC with vortices becomes a stationary state \cite{fetter} and the standard imaginary-time approach can be applied \cite{feder}. In the real-time approach \cite{ueda}, a dissipation has to be included in the GP equation to generate the vortices.
The imaginary-time approach \cite{feder} does not require any dissipation, is simpler to implement and is found to converge faster and lead to accurate results. Here we provide combined imaginary- and real-time programs in two (2D) and three (3D) spatial dimensions without any dissipation \cite{feder}. The present imaginary-time program already involves complex variables and is hence combined together with the real-time program. The choice of the type of propagation is made through an input parameter.
The imaginary-time approach should be used to solve the GP equation for the rotating BEC and to generate the stationary vortex lattice. A subsequent study of the non-stationary dynamics of the rotating BEC should be done using the real-time propagation. 
Here we provide C and Fortran programs for the
solution of the GP equation for a rotating BEC in a fully anisotropic 3D
trap by imaginary- and real-time propagation. We also present
C and Fortran programs for the reduced GP equation in 2D, appropriate for a disk-shaped BEC under a tight axial ($z$-direction) trapping.
We use the split-step Crank-Nicolson scheme for solving the GP equation, as in Refs.~\cite{bec2009,bec2012}. 
 
The imaginary-time algorithm employs a time iteration loop of an initial state until the convergence is reached \cite{bec2009}. 
The usual initial states are analytic wave functions, generally with one vortex at the center of the trap. However, such an analytic initial function may exhibit slow convergence and often may lead to an inappropriate final vortex lattice structure.
We will use an analytic initial function modulated by a random phase at different space points and 
show that this procedure is essential in addressing the convergence issues, as well as in obtaining the correct vortex lattice structure for a given set of system parameters.
Moreover, the GP equation of a rapidly rotating BEC with a very large number of vortices, viz. Figs.~\ref{fig2}(c) and (d) with 37 and 61 vortices, faces a convergence difficulty even after random phase modulation. In this latter case, when a pre-calculated converged wave function of the rotating BEC with a smaller number of vortices is used as the initial state, the convergence of the algorithm is vastly improved, resulting in the reduction of more than 90\% in execution time. 

In Section~\ref{sec:GPE} we present the GP equation for a rotating BEC in an anisotropic trap. We present the mean-field model
and a general scheme for its numerical solution. The reduced 2D GP equation appropriate for a disk-shaped
rotating BEC is also presented there.
The details about the computer programs, and their input/output
files, etc.~are given in Section~\ref{sec:details}. The numerical method and results
are given in Section~\ref{sec:numerics}, where we illustrate the generation of vortex lattices by employing the imaginary-time propagation in rapidly rotating trapped BECs with different angular frequencies and interaction strengths (nonlinearities). The stability of these vortex lattices is demonstrated in 
real-time propagation using the corresponding converged solution obtained by the imaginary-time propagation as initial states. The efficiency of parallelization of the present OpenMP programs in multi-core computers using the GNU and Intel compilers is also demonstrated there. 
Finally, a brief summary is given in Section~\ref{sec:con}.

\section{The Gross-Pitaevskii equation for a rotating condensate}
\label{sec:GPE}

A non-rotating BEC made up of $N$ atoms, each of mass $m$, can be described by the following mean-field GP equation for a wave function $\phi({\bf{r}},t)$ at the space point $\bf r$ at time $t$ \cite{fetter2}
\begin{equation}
{\mathrm i}\hbar \frac{\partial \phi({\bf r},t)}{\partial t}=\left[-\frac{\hbar^2}{2m}
\nabla^2_{\bf r} +\frac{1}{2}m\omega^2 (\gamma^2 x^2+\nu^2 y^2+\lambda^2 z^2)+\frac{4\pi \hbar^2 aN}{m}|\phi({\bf r},t)|^2
\right] \phi({\bf{r}},t)\, ,\quad \mathrm i=\sqrt{-1}\, ,
\end{equation}
where ${\bf r}\equiv ({\pmb{\rho}},z)\equiv ({x,y,z})$, $a$ is the atomic $s$-wave scattering length, and $\omega $ is the reference trapping frequency, with $\gamma, \nu, \lambda$ representing the trap anisotropies along the $x,y,z$ directions, respectively. The normalization condition is 
$\int d {\bf r} |\phi ({\bf{r}},t)|^2 =1$.
This equation can be derived from the energy functional \cite{fetter2}
\begin{equation}
E[\phi]=\int d {\bf r}\left[ \frac{\hbar^2}{2m}|\nabla _{\bf r}\phi|^2+ \frac{1}{2}m\omega^2 (\gamma^2 x^2+\nu^2 y^2+\lambda^2 z^2)|\phi|^2+ \frac{2\pi \hbar^2 aN}{m}|\phi|^4 \right]\, .
\end{equation}

The formation of a vortex lattice in a rapidly rotating BEC can be conveniently calculated 
in the rotating frame, where the generated vortex lattice forms a stationary state, which can be obtained by the imaginary-time propagation method. Such a dynamical equation in the rotating frame can be written if we note that the Hamiltonian in the rotating frame is
given by $ H = H_0 - \Omega L_ z$ \cite{landau}, where $ H _0$ is the laboratory frame Hamiltonian,
$\Omega$ is the angular frequency of rotation around the $z$ axis, and $ L_ z = \mathrm{i}\hbar(y\partial /\partial x - x\partial/\partial y)$ is the $z$ component of the angular momentum.
Consequently, the GP equation in the rotating frame 
has the explicit form \cite{fetter2,feder,zeng,ueda,jeng,bao}
\begin{eqnarray}
\mathrm i\hbar \frac{\partial \phi({\bf r},t)}{\partial t}=\left[-\frac{\hbar^2}{2m}\nabla^2_{\bf r} +\frac{1}{2} m\omega^2(\gamma^2 x^2+\nu^2 y^2+\lambda^2 z^2)+ \frac{4\pi \hbar^2 aN}{m} |\phi({\bf r},t)|^2
-\Omega L_z\right] \phi({\bf{r}},t)\, .
\end{eqnarray} 
Using the transformations ${\bf r}'={\bf r}/l,$ $ l=\sqrt{\hbar/(m\omega)}, a'=a/l$, $t'=\omega t$, 
$\phi'=l^{-3/2} \phi$, $\Omega'=\Omega/\omega$, and $ L_z '= L_z/\hbar$, we obtain the following convenient dimensionless form of the above equation
\begin{eqnarray}\label{3d}
\mathrm i \frac{\partial \phi({\bf r},t)}{\partial t}=\left[-\frac{1}{2}\nabla^2_{\bf r} +\frac{1}{2} (\gamma^2 x^2+\nu^2 y^2+\lambda^2 z^2)+ g_{3D} |\phi({\bf r},t)|^2
-\Omega L_z\right] \phi({\bf{r}},t)\, , \quad g_{3D} =4\pi N a\, ,
\end{eqnarray}
where we have dropped the primes from the transformed dimensionless variables.
We note that Eq.~(\ref{3d}) can also be derived from the dimensionless energy functional \cite{fetter2}
\begin{equation}\label{e3d}
E[\phi]=\int d {\bf r} \left[ \frac{1}{2}|\nabla_{\bf r} \phi|^2 + \frac{1}{2} (\gamma^2 x^2+\nu^2 y^2+\lambda^2 z^2)|\phi|^2+ \frac{1}{2}g_{3D} |\phi|^4 - \phi^* \Omega L_z \phi \right]\, ,
\end{equation}
obtained using the same transformations and expressing the energy in units of $\hbar\omega$.
All derivations and results presented in the following are using these dimensionless variables. 

A convenient equation for a quasi-2D disk-shaped BEC under a strong harmonic confinement in the $z$ direction ($\lambda >>\gamma, \nu $) can be derived using the following ansatz for the wave function \cite{luca}
\begin{eqnarray}\label{anz}
\phi({\bf r},t)= \psi ({\pmb \rho},t)\times \frac{1}{(\pi d_z^2)^{1/4}}\exp\left(-\frac{z^2}{2d_z^2}\right)\, , \quad d_z= \sqrt{\frac{1}{\lambda}}\, ,
\end{eqnarray} 
where we assume that because of the strong confinement the dynamics in the $z$ direction will be frozen to a time-independent Gaussian of width $d_z$, and that the relevant dynamics will evolve only in the $x$-$y$ plane. If we substitute the ansatz 
(\ref{anz}) to Eq.~(\ref{3d}), we can integrate out the $z$ variable and obtain the 
corresponding dynamical equation in 2D, valid for a quasi-2D rotating BEC in a disk-shaped trap \cite{bec2009,luca}:
\begin{eqnarray}\label{2d}
\mathrm i\frac{\partial \psi({\pmb \rho},t)}{\partial t}=\left[-\frac{1}{2}\nabla^2_{\pmb \rho} +\frac{1}{2} (\gamma^2 x^2+\nu^2 y^2)+g_{2D}|\psi({\pmb \rho},t)|^2
-\Omega L_z\right] \psi({\pmb \rho},t)\, , \quad g_{2D}=\frac{4\pi a N \sqrt \lambda}{\sqrt{2\pi}}\, ,
\end{eqnarray} 
with the normalization condition $\int d {\pmb \rho} |\psi({\pmb \rho},t)|^2 =1$.
The energy functional corresponding to Eq.~(\ref{2d}) is
 \begin{eqnarray}\label{e2d}
E[\psi]=\int d {\pmb \rho}\left[ \frac{1}{2} |\nabla_{\pmb \rho} \psi|^2+ \frac{1}{2} (\gamma^2 x^2+\nu^2 y^2)|\psi|^2+ \frac{1}{2}g_{2D}|\psi|^4 - \psi^* \Omega L_z \psi \right].
\end{eqnarray} 

We use the split-step Crank-Nicolson algorithm for the solution of the GP equations (\ref{3d}) and (\ref{2d}). This approach has been elaborated in details in Ref.~\cite{bec2009}. In the following we describe the necessary modifications for the 2D equation (\ref{2d}). We follow the identical prescription in 3D. Noting that 
$L_z=\mathrm{i}\hbar(y\partial /\partial x - x\partial/\partial y)$, we split the Hamiltonian into three parts: 
\begin{eqnarray}
H&\equiv &H_1+H_2+H_3\, ,\\
H_1&=& \frac{1}{2}(\gamma^2 x^2+\nu^2 y^2)+g_{2D}|\psi|^2, \label{eq1}\\
H_2&=& - \frac{1}{2} \frac{\partial }{\partial x^2} - {\mathrm i} \Omega y \frac{\partial}{\partial x}\, , \label{eq2}\\
H_3&=& - \frac{1}{2} \frac{\partial }{\partial y^2} + {\mathrm i} \Omega x \frac{\partial}{\partial y}\, . \label{eq3}
\end{eqnarray}
In this approach we perform the time propagation over infinitesimally small time step first over only the part $H_1$, and then over the part $H_2$, and finally over the part $H_3$ of the Hamiltonian. Essentially, we split Eq.~(\ref{2d}) into:
\begin{eqnarray}\label{split}
\mathrm i\frac{\partial \psi}{\partial t}= H_1 \psi, \quad \mathrm i\frac{\partial \psi}{\partial t}= H_2 \psi, \quad \mathrm i\frac{\partial \psi}{\partial t}= H_3 \psi,
\end{eqnarray}
and perform the time propagation over these three sub-equations successively and independently of each other, in the given order.

We first solve the first of Eqs.~(\ref{split}) 
starting from an initial state $\psi (\pmb \rho, t_0 ) $ at $t = t_0$ to obtain the first intermediate solution after an infinitesimal time step $\Delta$. 
Then this intermediate solution is used as an initial value to solve the second of Eqs.~(\ref{split}), yielding the second intermediate solution at the time $t=t_0+\Delta$, which is then used to propagate the third of Eqs.~(\ref{split}) over the infinitesimal time $\Delta$ to yield the final solution at $t=t_0+\Delta$, after one full time iteration of Eq.~(\ref{2d}). This
procedure is repeated $n$ times to get the final solution at time $t_{\mathrm{ final}} = t_ 0 + n \Delta$.
 
The first equation of (\ref{split}) with $H_1$ has the analytic solution \cite{bec2009}, which we denote by $\psi^{k+1/3}$ when propagating between the time steps $k$ and $k+1$. Similarly, we denote by $\psi^{k+2/3}$ the wave function after the time propagation with respect to $H_2$, and finally by $\psi^{k+1}$ after additional propagation with respect to $H_3$, i.e., after one full time iteration. Following Ref.~\cite{bec2009} and using notations therein, we discretize the second equation of (\ref{split}) for $H_2$ alone as 
\begin{align}\label{eqD.9:kn1}
\mathrm{i}\frac{\psi_{i}^{k+2/3}-\psi_{i}^{k+1/3}}{\Delta} = & -{\frac{1}{2}}\frac{1}{2h_x^2}\biggr\{\left(\psi^{k+2/3}_{i+1}-2\psi^{k+2/3}_{i} +\psi^{k+2/3}_{i-1}\right)
+\left(\psi^{k+1/3}_{i+1}-2\psi_{i}^{k+1/3}+\psi^{k+1/3}_{i-1}\right)\biggr\} \notag \\ 
& -\frac{{\mathrm i}\Omega y_j}{4 h_x}\biggr\{\left(\psi^{k+2/3}_{i+1}-\psi^{k+2/3}_{i-1}\right) + \left(\psi^{k+1/3}_{i+1}-\psi^{k+1/3}_{i-1}\right) \biggr\} ,
\end{align}
where $\psi_{i}^t=\psi(x_i, y_j, t)$ refers to the wave function value at the spatial grid point determined by $x\equiv x_i=-N_xh_x/2+ ih_x, y_j= -N_yh_y/2+ jh_y $, $i=0,1,2,\ldots,N_x$, and $j=0,1,2,\ldots,N_y$. Here $h_x,h_y$ are the space steps along the $x$ and $y$ directions, respectively, and $t=k+1/3$ or $k+2/3$ refers to the time iteration \cite{bec2009}, connecting the present $(k+1/3)$ to the future $(k+2/3)$ in propagation with respect to $H_2$.

The above procedure results in a set of tridiagonal equations (\ref{eqD.9:kn1}) in $\psi^{k+2/3}_{i+1}$, $\psi^{k+2/3}_{i}$, and $\psi^{k+2/3}_{i-1}$ at time $t_{k+2/3}$, which are solved using the proper boundary conditions \cite{bec2009}. The tridiagonal equations are written explicitly as $A_i^-\psi^{k+2/3}_{i-1}+A_i^0\psi^{k+2/3}_{i}+
A_i^+\psi^{k+2/3}_{i+1}= b_i,$ where
\begin{align}\label{eqD.11}
b_i={\frac{{\mathrm i \Delta}}{4h_x^2}}
\left(\psi^{k+1/3}_{i+1}-2\psi_{i}^{k+1/3}+\psi^{k+1/3}_{i-1}\right) 
-\frac{{ \Delta} { \Omega} y_j}{4h_x}\left(\psi^{k+1/3}_{i+1}-\psi^{k+1/3}_{i-1}\right)+\psi_{i}^{k+1/3}, \\
A_i^0 =1+ \frac{{\mathrm i \Delta}}{{2}h_x^2}, \;\;\; A_i^-= -\frac{ {\mathrm i \Delta}}{{4}h_x} \left( \frac{1}{h_x}-{\mathrm i\Omega y_j} \right), \;\;\; A_i^+=- \frac{{\mathrm i \Delta}}{{4}h_x} \left( \frac{1}{h_x}+{ \mathrm i \Omega y_j} \right). \label{eq5}
\end{align}
The discretization for $H_3$ is performed similarly. 
The tridiagonal set of equations above is very similar to Eqs.~(34) and (35) of Ref.~\cite{bec2009}, and the real-time propagation routine is programmed and solved in identical fashion after a straightforward modification to include the extra terms due to a non-zero value of $\Omega$ in these equations. The imaginary-time propagation routine corresponds to a transformation $t \to -\mathrm i t$ or $\Delta \to -\mathrm i \Delta$ \cite{bec2009}
and hence can be obtained by replacing $\mathrm i\Delta \to \Delta$ in Eqs.~(\ref{eqD.11}) and (\ref{eq5}) in the real-rime routine, which is performed in our combined real- and imaginary-time programs by the selection parameter \texttt{OPTION\_RE\_IM}.

Instead of evaluating the real energies from Eqs.~(\ref{e3d}) and (\ref{e2d}) in 3D and 2D 
involving complex algebra over complex wave functions, it is convenient to write a real expression for the energy. To calculate the energy and the chemical potential, we write the two coupled nonlinear equations for the real and imaginary parts of the wave function ($\psi=\psi_R+\mathrm i \psi_I$), viz.~Eqs.~(2.1) of Ref.~\cite{jeng}. The equation satisfied by the real part is 
 \begin{align}\label{real}
\mathrm i\frac{\partial \psi_R({x,y};t)}{\partial t}=\left[-\frac{1}{2}\nabla^2 +\frac{1}{2} (\gamma^2 x^2+\nu^2 y^2)+g_{2D}|\psi({x,y};t)|^2\right] \psi_R({x,y};t)
+\Omega\left( y\frac{\partial} {\partial x} -x \frac{\partial}{\partial y} \right) \psi_I({x,y};t)\, .
\end{align}
In this equation $\psi_R$ is not normalized to unity.
Using Eq.~(\ref{real}), the energy and the chemical potential can be expressed in 2D as 
\begin{align}\label{e2dr}
\frac{1}{{\int dx dy\, \psi_R^2 }}{\int d{\pmb \rho} \left[-{\frac{1}{2}{(\nabla_{\pmb \rho} \psi_R)^2}}+ \frac{1}{2}(\gamma^2 x^2+ \nu^2 y^2) \psi_R^2+ \alpha g_{2D}(\psi_R^2+\psi_I^2){\psi_R^2} +\Omega \psi_R \left( y\frac{\partial }{\partial x} -x \frac{\partial}{\partial y} \right)\psi_I\right] },
\end{align}
where the value $\alpha=1 $ corresponds to the chemical potential $\mu$, and the value $\alpha=1/2$ to the energy $E$ per atom. A similar expression for energy and chemical potential in 3D is 
\begin{align}\label{e3dr}
\frac{1}{{\int d{\bf r}\, \phi_R^2 }}{\int d{\bf r} \left[-{\frac{1}{2}{(\nabla_{\bf r} \phi_R)^2}}+ \frac{1}{2}(\gamma^2 x^2+ \nu^2 y^2+\lambda^2 z^2) \phi_R^2+\alpha g_{3D}(\phi_R^2+\phi_I^2){\phi_R^2} +\Omega \phi_R \left( y\frac{\partial }{\partial x} -x \frac{\partial}{\partial y} \right)\phi_I\right] }.
\end{align}
Equations~(\ref{e2dr}) and (\ref{e3dr}) are equivalent to Eqs.~(\ref{e2d}) and (\ref{e3d}) and involve algebra of real functions. Hence these equations lead to far more accurate numerical results than the previous set of expressions. Specifically, the calculation of the rotational energy and the kinetic energy term involving derivatives and gradients of a complex wave function in Eqs.~(\ref{e2d}) and (\ref{e3d}) can be numerically problematic. 

The initial wave function in the imaginary-time programs is taken to be {one} containing a single vortex at the center, aligned with the $z$ axis. Explicitly, for the 2D and 3D programs, we take, respectively
\begin{equation} \label{fun}
\psi_{\mathrm{initial}}(x,y) = \frac{{x+\mathrm i y}}{\sqrt {\pi d_{xy}^2}}\exp \left( -\frac{x^2+y^2}{2d_{xy}^2}{+2\pi \mathrm i {\cal R}(x,y)}\right), \quad
\phi_{\mathrm{initial}}(x,y,z)=\psi_{\mathrm{initial}}(x,y) \frac{1}{(\pi d_z^2)^{1/4}}
\exp \left( -\frac{z^2}{2d_z^2}\right),
\end{equation}
where $d_{xy}$ and $d_z$ are width parameters in the $x$-$y$ plane and in the $z$ direction, {and ${\cal R}(x,y)$ is a random number. In numerical calculation, the random phase ensures that the number of vortices changes by units of one, as parameters, e.g., nonlinearity and angular frequency, are changed. Without the random phase, the number of vortices changes by units of two or multiples of two. In fact, any localized normalizable initial function modulated by a random phase at different space points, e.g., a Gaussian function without any vortices, obtained by setting $(x+\mathrm i y)=1$ in Eq.~(\ref{fun}), will lead to the same vortex lattice as the initial function (\ref{fun}) with one vortex. Without the random phase these functions usually will lead to different results \cite{bao}.}

\section{Details about the programs}
\label{sec:details}

All input data (number of atoms, scattering length, harmonic oscillator trap length, trap anisotropy, etc.) are conveniently placed at the beginning of each Fortran program, as before \cite{bec2016}. Hence after changing the input data in a Fortran program a recompilation is required. The C programs use external input files that contain all parameters, and their adjustment does not require a recompilation. 
The source programs are located in the directory \texttt{src} within the corresponding package directory (\texttt{BEC-GP-ROT-OMP-C} for the C programs and \texttt{BEC-GP-ROT-OMP-F} for the Fortran ones). They can be compiled by the \texttt{make} command using the \texttt{makefile} in the corresponding package root directory. The examples of produced output files can be found in the directory \texttt{output}, although some large density files are omitted, to save space. { The programs use an initial state with repeatable random phase. A different random phase can be generated by changing the variable SEED in the subroutine Initialize for the Fortran programs, or in the corresponding input file for the C programs.
The provided Fortran output files are calculated with SEED = 13 using the one-vortex initial function (\ref{fun}). The change of the variable SEED implies a different initial function, thus changing the output files. In the Fortran programs, the random phase is included by the integer parameter RANDOM: the value 0 excludes the random phase and 1 includes it. The integer parameter FUNCTION permits the selection of a Gaussian or a one-vortex initial function: the value 0 selects a Gaussian function and 1 selects the one-vortex function (\ref{fun}). For the C programs, the input files contain variables providing the same functionality, which is explained there.}
After running a program and obtaining the results, one can use the file \texttt{fig*.gnu} in the directory \texttt{output} to visualize the density profiles, relying on a popular software package \texttt{gnuplot}. These files are used by invoking the command \texttt{gnuplot fig*.gnu} to obtain an eps figure of the generated vortex lattice. Depending on the density file to be plotted, one has to adjust the corresponding line in the \texttt{fig*.gnu} file. Currently it is set to use the density file provided as an example and already present in the \texttt{BEC-GP-ROT-OMP} distribution.

The output files are conveniently named such that their contents can be easily identified, following the naming convention introduced in Ref.~\cite{bec2016}. For example, a file named \texttt{<code>-out.txt}, where \texttt{<code>} is a name of the individual program, represents the general output file containing input data, time and space steps, nonlinearity, energy, and chemical potential. A file named \texttt{<code>-den2d.txt} is the output file with the reduced (integrated) {2D} condensate density.
There are output files for reduced (integrated) 1D densities for different programs. Typically, a user first solves the stationary problem using the imaginary-time programs, and then uses the real-time programs to read the pre-calculated stationary wave function and to study the dynamics. To read the pre-calculated wave function
 the parameter \texttt{NSTP} should be set to zero. In this way one can also run the imaginary time program with a pre-calculated wave function. The supplied programs have the pre-defined value \texttt{NSTP = 1} and use the analytic wave function (\ref{fun}) as the initial state. 
In each program the selection for imaginary- or real-time propagation is done by setting the parameter \texttt{OPTION\_RE\_IM} to 1 or 2, respectively. If the imaginary-time propagation is thus selected, the programs run either by using an initial 
analytic input function (if \texttt{NSTP} is not set to zero) or by employing a pre-calculated wave function (if \texttt{NSTP} is set to zero). The real-time propagation can successfully work only with a meaningful initial wave function, usually assuming that \texttt{NSTP = 0} is set, and that the program will read a pre-calculated wave function by the earlier performed imaginary-time propagation. 
The reader is advised to consult our previous publication where a complete description of the output files is given \cite{bec2017x}. The calculation is essentially done in the \texttt{NPAS} time loop, which are in the Fortran programs conveniently divided into 10 equal intervals (\texttt{NPAS/10}). The output files for the reduced 2D densities at the end of each of these intervals are saved as files \texttt{<code>*-den-j.txt}, where \texttt{j}=1,\ldots,10. If necessary, one can further customize this by changing and recompiling the Fortran programs. In the C programs the selection of output files is done through the input file, when one can set the desired frequency of saving the output densities, as well as the types of density profiles to be saved. A \texttt{README.md} file, included in the corresponding root directory for C and Fortran, explains the procedure to compile and run the programs in more detail.

The supplied 2D programs are preset to run the imaginary-time propagation using the space steps DX=DY =0.05, numbers of space points NX=NY=256, $g_{2D}=100, \Omega =0.8$, the trap parameters $ \gamma =\nu = 1$. The 3D programs use DX=DY =0.05, DZ=0.025, NX=NY=256, NZ=32,
$\gamma=\nu=1, \lambda =100, g_{2D}=100, g_{3D}=g_{2D}\sqrt{2\pi/\lambda}={25.0662827} $. 
The large trap parameter $\lambda$ ensures a disk-shaped BEC, which enables a comparison of the 3D results for the integrated density over the $z$ coordinate with the 2D density profile. This also reduces a transversal instability of the 3D vortex lines. The time steps used are $\Delta=0.00025$ (imaginary time) and 0.0001 (real time), numbers of time iterations are NPAS=3,000,000 and NSTP=1 (imaginary time) and NSTP=0 (to run real- or imaginary-time propagation with a pre-calculated wave function as an input). To achieve the convergence in some cases (large nonlinearity $g_{2D}, g_{3D}$ and $\Omega$), one may need to increase the values of NX, NY, NPAS, and reduce the space and time steps DX, DY, DZ and DT accordingly. Note that the actual spatial grid used contains (NX+1)$\times$(NY+1) or (NX+1)$\times$(NY+1)$\times$(NZ+1) points, since in each dimension the grid index takes the values from 0 to NX, etc. Therefore, the produced output files also contain the data for such grid sizes.

The function (\ref{fun}) always leads to a converged solution after a large number of time iterations in imaginary-time propagation.
{A Gaussian wave function given as an input in imaginary-time propagation would sometimes face a convergence difficulty and should not be used.} Therefore the programs by default use a better initial state, containing one vortex at the center. Once a stationary vortex lattice is obtained for a specific nonlinearity and angular frequency by imaginary-time propagation, the final wave function so obtained should be used as the initial state for the generation of vortex lattices by imaginary-time propagation with larger nonlinearities and/or angular frequencies. For example, 
to generate closed hexagonal vortex lattices of 19, 37, and 61 vortices in the panels (b), (c) and (d) of Figs.~\ref{fig2} and \ref{fig5} in the next section, respectively, we have used the previously calculated initial states of 7, 19, and 37 vortices in the corresponding panels (a), (b), and (c), respectively.
Such a choice of dynamically generated multi-vortex initial state with a proper phase distribution enhances the convergence of the numerical scheme enormously compared to the propagation starting from a single-vortex initial state. The reduction in the execution time for the calculation done in this fashion could be as much as 99\%. The size of the condensate increases as the nonlinearity and/or the angular frequency $\Omega$ are increased. To accommodate a larger condensate, the number of space points NX, NY, etc. should be appropriately increased. To read a pre-calculated wave function by setting NSTP to zero, the grid size in the used wave function file should match exactly the number of points used in the current program. The supplied programs assume equal numbers of space step points in both imaginary- and real-time propagation, and in C programs this is configurable through the input files. If the grid sizes in the two calculations are different, the user can customize the programs to accommodate this. For instance, in Fortran programs the READ statement in the subroutine INITIALIZE should be changed, for instance, from \texttt{I=0,NX} to \texttt{I= NX2-NXOLD2,NX2+NXOLD2,1}, where \texttt{NXOLD2} is the \texttt{NX2} value of the previous calculation with a smaller number of grid points.

\section{Numerical results}
\label{sec:numerics}

\begin{figure}[!ht]
\centering
\includegraphics[width=.7\linewidth,clip]{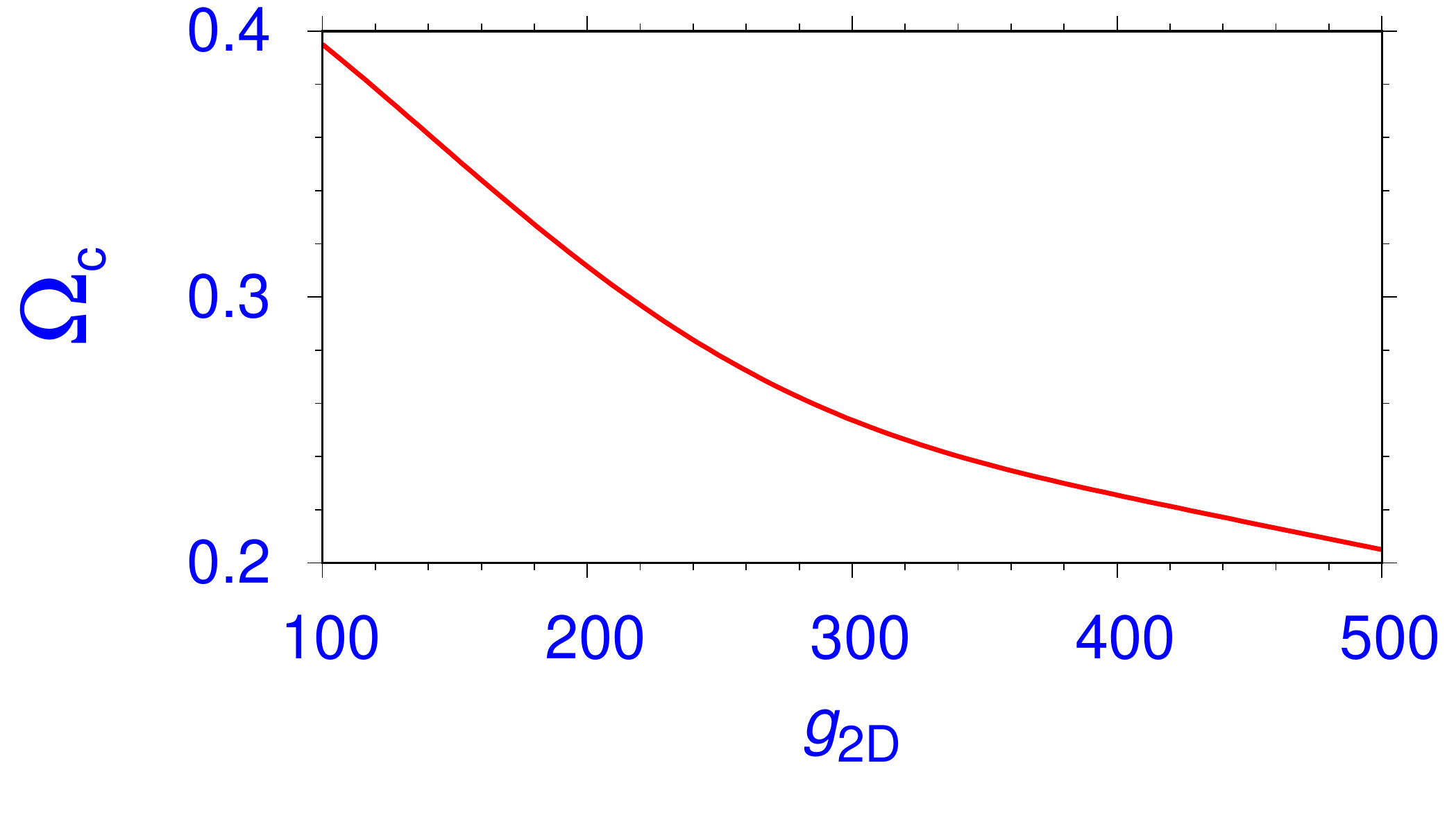}
\caption{{ Critical angular frequency $\Omega_c$ for the generation of a single vortex using function (\ref{fun}) with random phase versus nonlinearity $g_{2D}$ for a rotating BEC in 2D. For $\Omega < \Omega_c$ no vortex is generated. }
}
\label{fig1}
\end{figure}

To test the programs and to demonstrate their usage, we have generated vortex-lattice structures using the imagi\-na\-ry-time programs and then ran the real-time programs starting from the previously obtained imagi\-na\-ry-time wave functions as inputs. {First, we numerically calculate the critical angular frequency $\Omega_c$, for the generation of a single vortex, using the initial function (\ref{fun}), for a rotating BEC in 2D for different nonlinearities $g_{2D}$. Without the random phase in the initial wave function this threshold cannot be calculated, as, then, a single vortex continues to exist for $\Omega < \Omega_c$. The result is displayed in Fig.~\ref{fig1}. The displayed result is the average over several runs.}

\begin{figure}[!t]
\centering
\includegraphics[height=.195\linewidth]{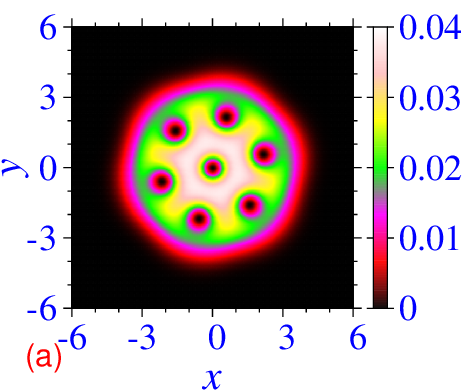}\hspace*{1mm}
\includegraphics[height=.195\linewidth]{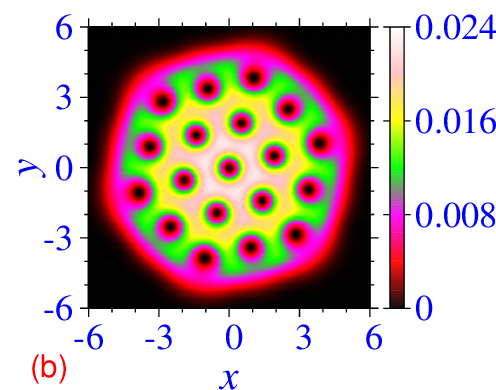}\hspace*{1mm}
\includegraphics[height=.195\linewidth]{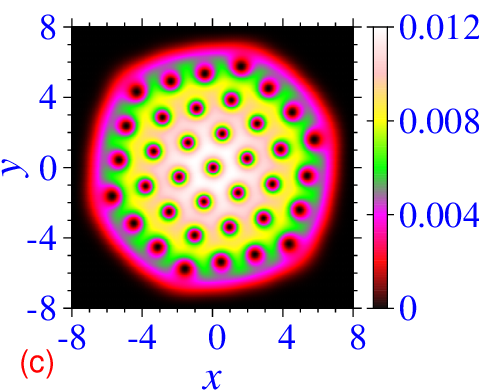}\hspace*{1mm}
\includegraphics[height=.195\linewidth]{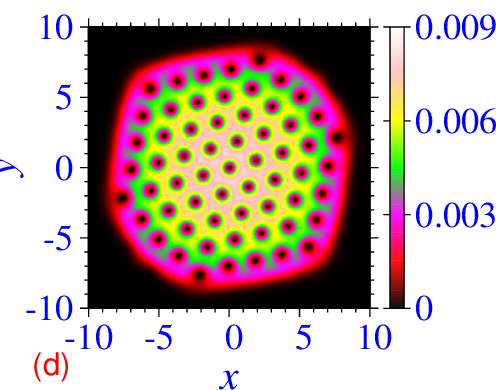}\vspace*{1mm}
\includegraphics[height=.195\linewidth]{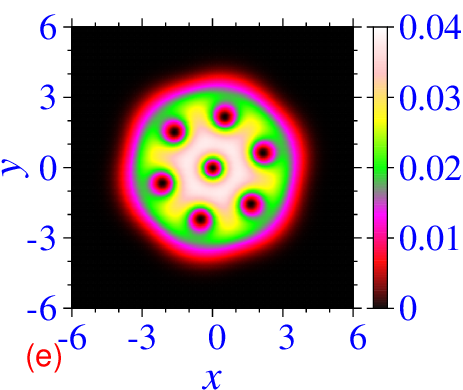}\hspace*{1mm}
\includegraphics[height=.195\linewidth]{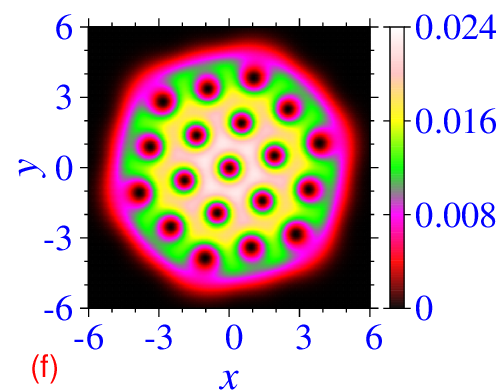}\hspace*{1mm}
\includegraphics[height=.195\linewidth]{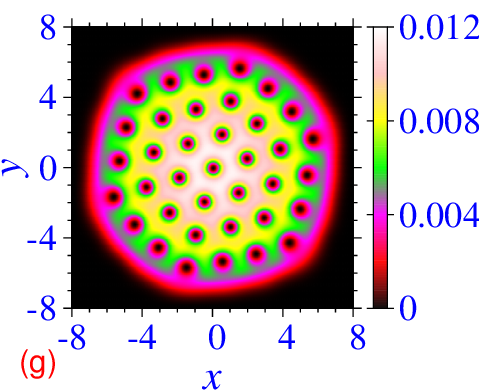}\hspace*{1mm}
\includegraphics[height=.195\linewidth]{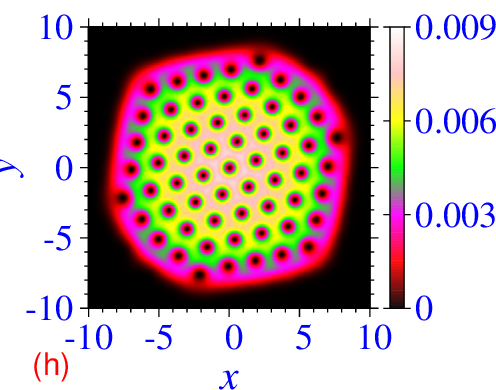}
\caption{Contour plots of the density $|\psi(x,y)|^2$ for the generated vortex lattices by the 2D imaginary-time propagation of Eq.~(\ref{2d}) for: (a) $g_{2D}=100$, $\Omega = 0.8$, (b) $g_{2D}=100$, $\Omega = 0.95$, (c) $g_{2D}=500$, $\Omega = 0.92$, (d) $g_{2D}=500$, $\Omega = 0.978$. Panels (e), (f), (g), and (h) display these vortex lattices, respectively, after the additional real-time propagation for 500 units of time using the corresponding imaginary-time wave function as input. The employed trap parameters are $ \nu=\gamma=1$, the space steps are DX=DY=0.05, and the time steps are 0.00025 in imaginary time and 0.0001 in real time. The size of the condensate increases as $\Omega$ increases from (a) to (b) and from (c) to (d), and as $g_{2D}$ increases from (b) to (c). The space grids used are: (a) $257\times 257$, (b) $321\times 321$, (c) $401\times 401$, and (d) $441\times 441$.
}
\label{fig2}
\end{figure}

\begin{figure}[!t] 
\centering
\includegraphics[height=.195\linewidth]{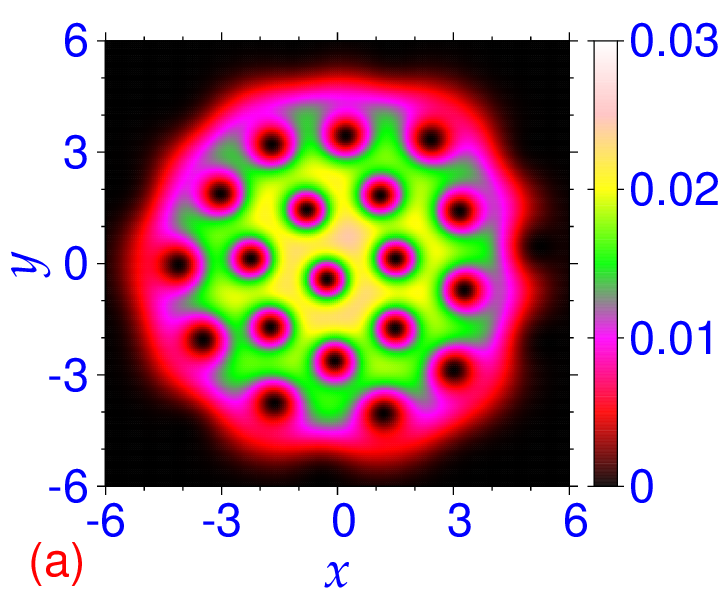}\hspace*{1mm}
\includegraphics[height=.195\linewidth]{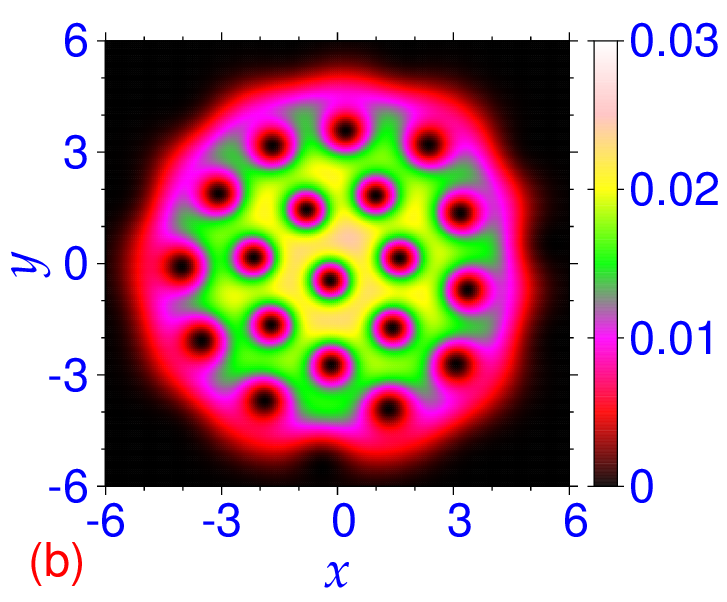}\hspace*{1mm}
\includegraphics[height=.195\linewidth]{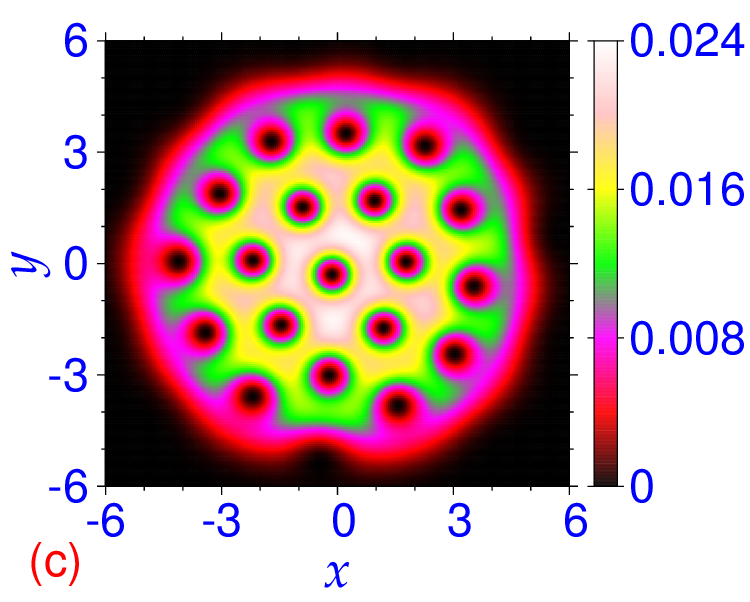}\hspace*{1mm}
\includegraphics[height=.195\linewidth]{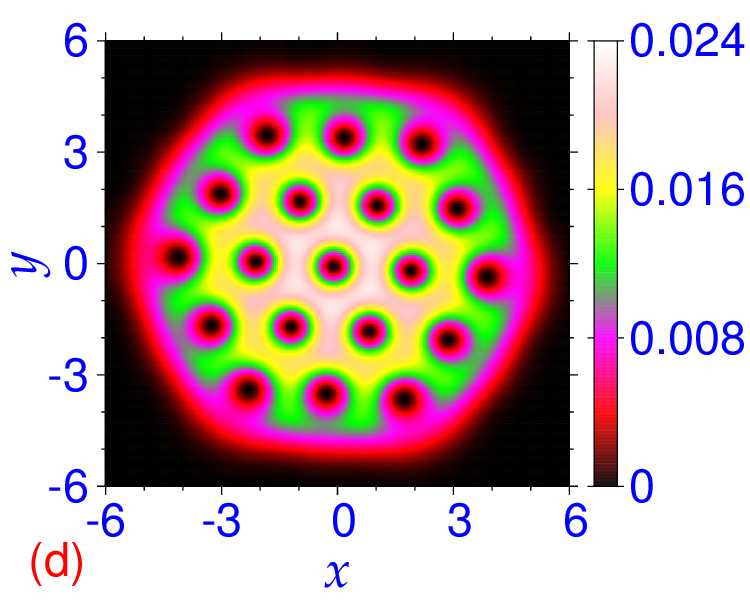}\vspace*{1mm}
\includegraphics[height=.2\linewidth]{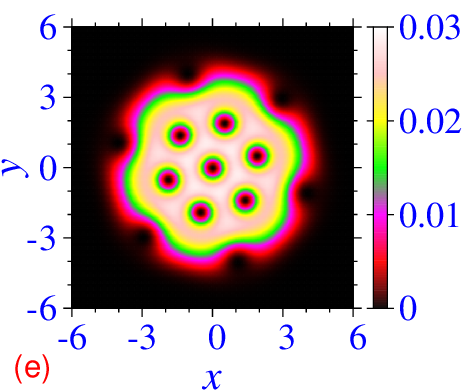}\hspace*{1mm}
\includegraphics[height=.2\linewidth]{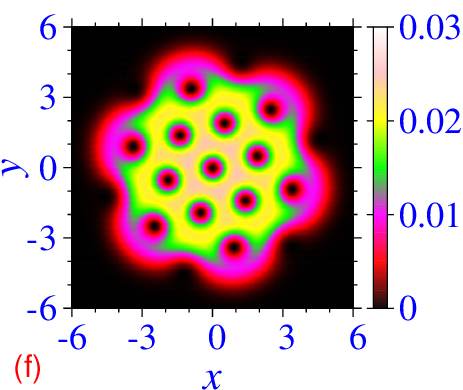}\hspace*{1mm}
\includegraphics[height=.2\linewidth]{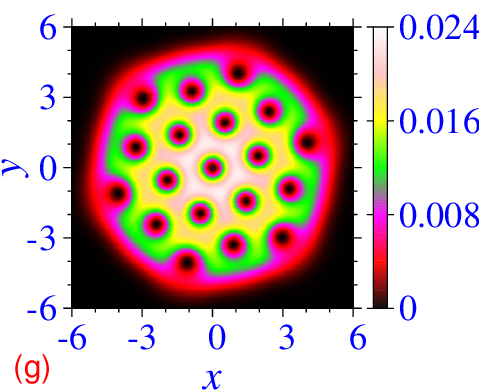}\hspace*{1mm}
\includegraphics[height=.2\linewidth]{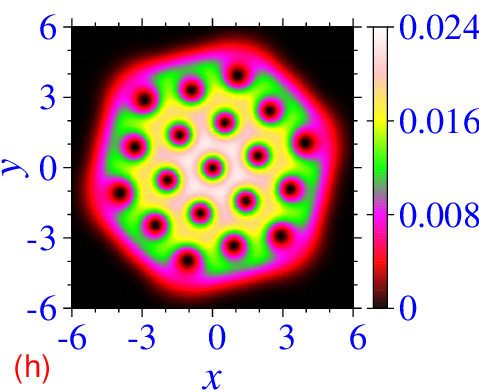}
\caption{The convergence of calculation from snapshots at different time steps during the imaginary-time propagation of the 2D equation (\ref{2d}) to generate a vortex lattice for parameters $g_{2D}=100$, $\Omega=0.95$. Numerical simulation used the initial state (\ref{fun}), and the panels correspond to: {(a) $2 \times 10^5$, (b) $4 \times 10^5$, (c) $8\times 10^5$, (d) $12\times 10^5$} time steps. For the same parameters, a much faster convergence is obtained in a simulation using as the initial function the converged wave function from Fig.~\ref{fig2}(a), obtained for $g_{2D}=100$, $\Omega=0.8$. The panels correspond to: (e) $5,000$, (f) $10,000$, (g) $20,000$, (h) $ 30,000$ time steps. 
The employed time step is 0.00025, the space steps DX=DY=0.05, and the grid size used is $321\times 321$ in all panels.
}
\label{fig3}
\end{figure}

We {next} numerically study the 2D vortex lattice in a rotating BEC using the imaginary-time propagation. 
The imaginary-time propagation with the 
supplied 2D program bec-gp-rot-2d-th uses the 
wave function (\ref{fun}) as the initial state and the parameters $g_{2D}=100$ and $ \Omega=0.8$. The generated
vortex lattice with seven vortices arranged in a triangular lattice in the shape of a closed hexagon is exhibited in Fig.~\ref{fig2}(a) through the contour density plot. 
In Fig.~\ref{fig2}(b) we illustrate the 2D vortex lattice with 19 vortices arranged in a triangular lattice in the shape of a closed hexagonal form obtained with parameters $g_{2D}=100$, $\Omega = 0.95.$ To illustrate the convergence of the imaginary-time propagation we show in Figs.~\ref{fig3}(a)-(d) the 2D density profiles at different times, using the analytic wave function (\ref{fun}) as the initial state and employing the parameters $g_{2D}=100$, $\Omega = 0.95$, the same as in Fig.~\ref{fig2}(b). This scheme shows a { slow} convergence and the vortex lattice structure practically remains the same from the panel \ref{fig3}(a) for {$2\times 10^5$} time steps to the panel \ref{fig3}(c) for 
{$8\times 10^5$} time steps with {19} vortices, before converging to the desired solution in the panel \ref{fig3}(d) after {$12\times 10^5$ } time steps, containing 19 vortices.
The convergence can be highly enhanced if we use the final converged state with a smaller number of vortices as the initial state of a calculation where a larger number of vortices is expected, either because the parameters $g_{2D}$ or $\Omega$ or both 
are larger. In Fig.~\ref{fig3}(e)-(h) we demonstrate this and show the vortex lattice evolution of the rotating BEC for the same parameters $g_{2D}=100$, $\Omega = 0.95$ as in the panels \ref{fig3}(a)-(d), but starting from the initial state with seven vortices, obtained in Fig.~\ref{fig2}(a) for $g_{2D}=100$, $\Omega = 0.8$. In Fig.~\ref{fig3} we see that the convergence in this case is achieved much faster. In practical terms, in panels \ref{fig3}(c) after 20,000 time steps or \ref{fig3}(d) after 30,000 time steps of the imaginary-time propagation the convergence is already reached. The reduction in execution time in the later scheme resulting in Figs.~\ref{fig3}(e)-(h) 
compared to the former resulting in Figs.~\ref{fig3}(a)-(d) could be { very large, viz.~$12\times 10^5$} time iterations and 30,000 time iterations in the two schemes.

{
\begin{figure}[!t] 
\centering
\includegraphics[height=.18\linewidth]{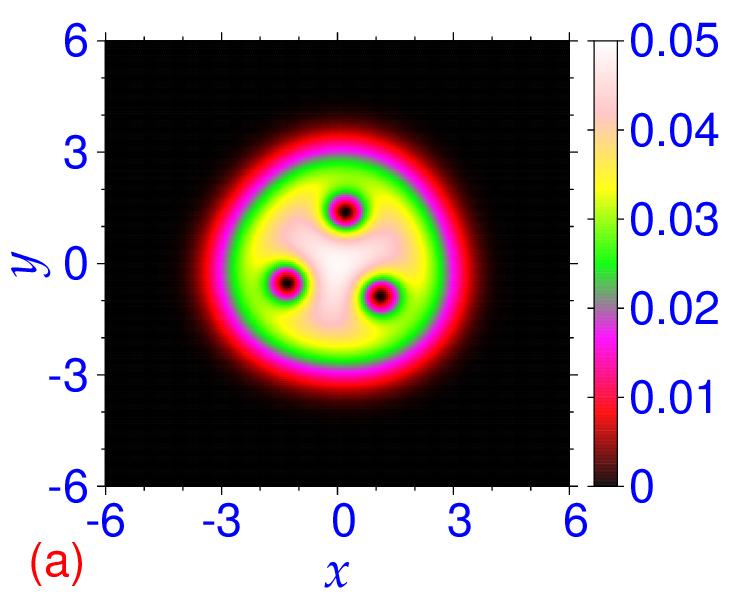}\hspace*{1mm}
\includegraphics[height=.18\linewidth]{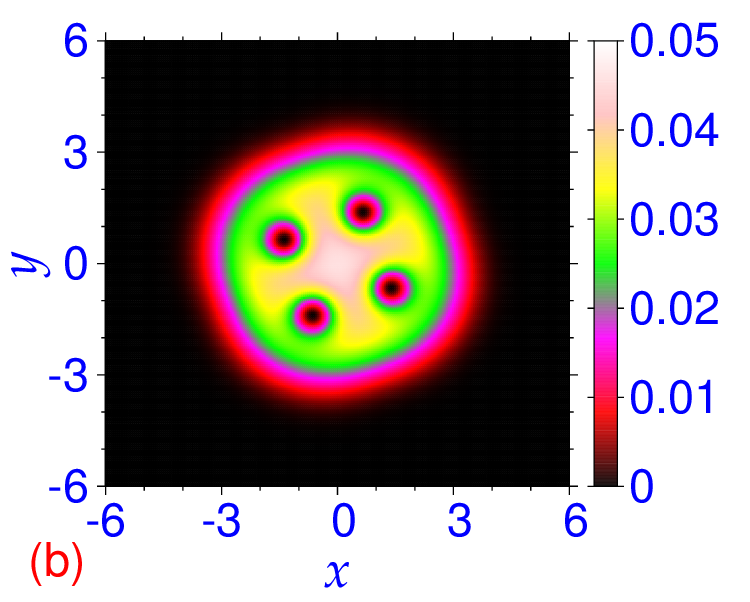}\hspace*{1mm}
\includegraphics[height=.18\linewidth]{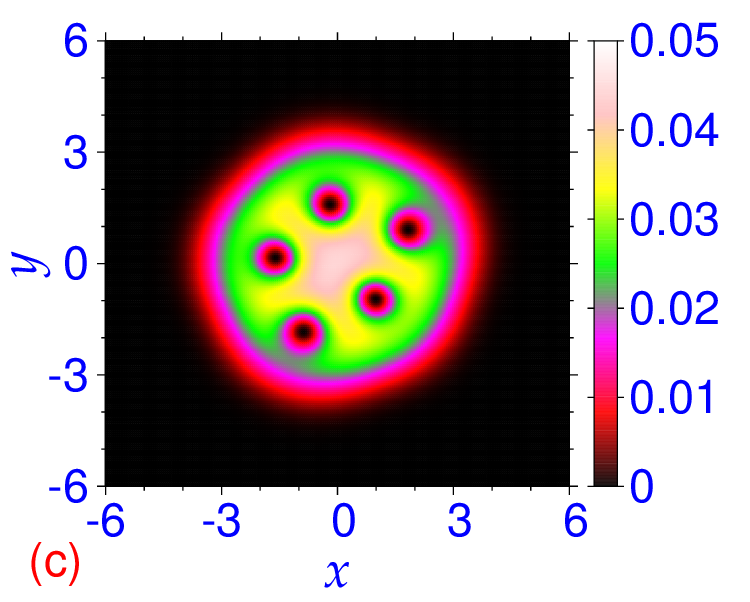}\hspace*{1mm}
\includegraphics[height=.18\linewidth]{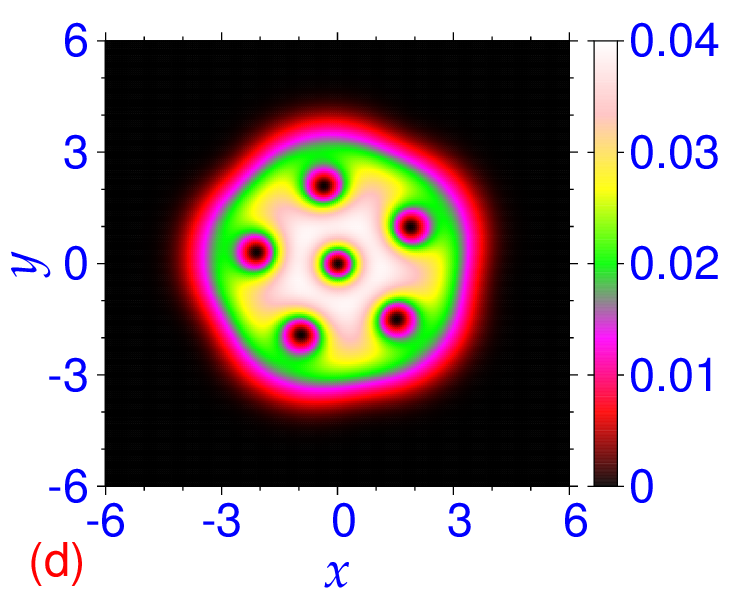}\vspace*{1mm}
\includegraphics[height=.18\linewidth]{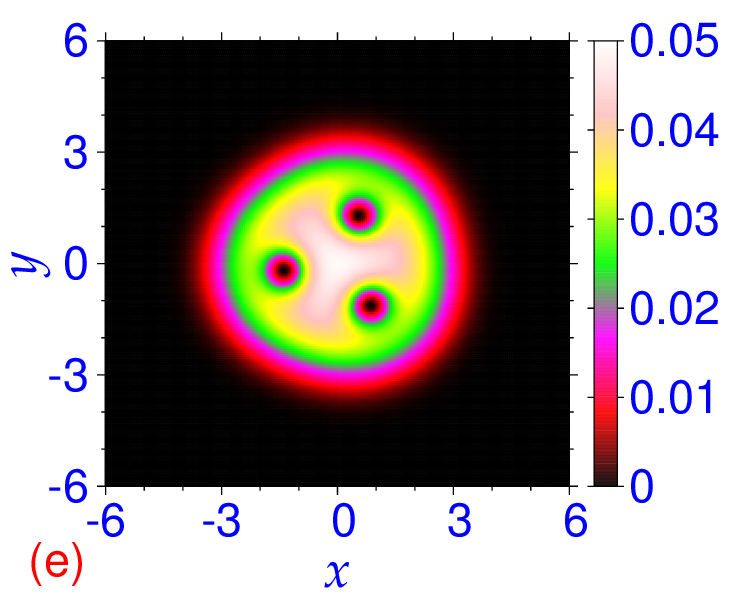}\hspace*{1mm}
\includegraphics[height=.18\linewidth]{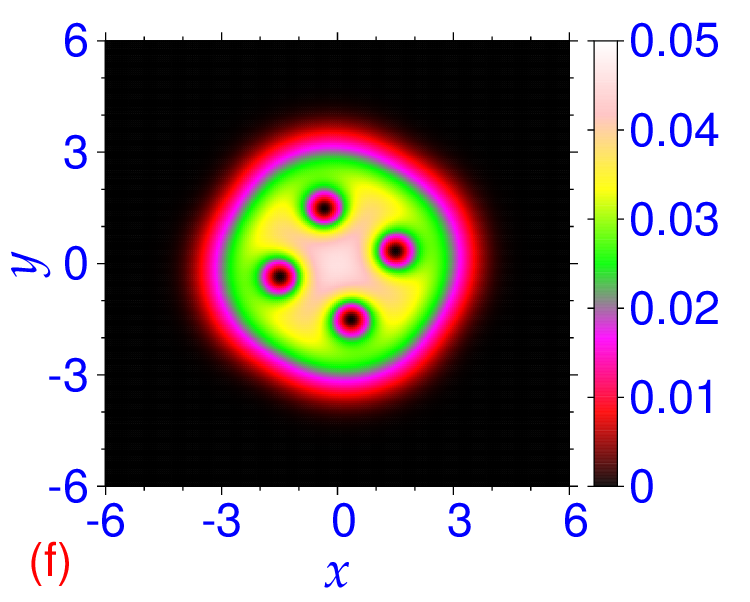}\hspace*{1mm}
\includegraphics[height=.18\linewidth]{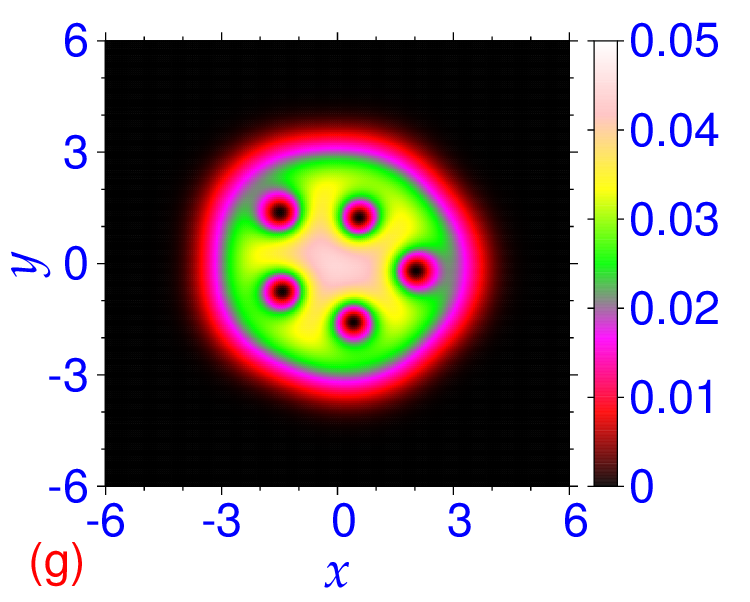}\hspace*{1mm}
\includegraphics[height=.18\linewidth]{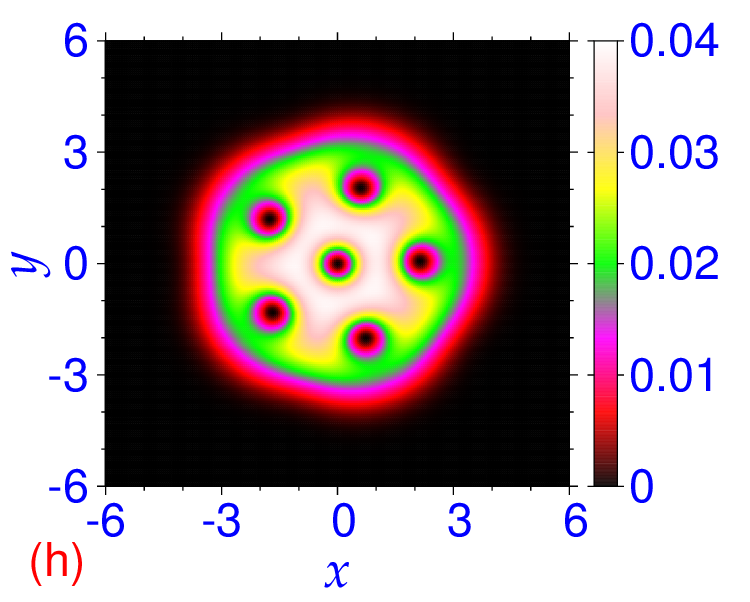} 
\caption{Contour plots of the density $|\psi(x,y)|^2$ for the generated vortex lattices by the 2D imaginary-time propagation of Eq.~(\ref{2d}) for $g_{2D}=100$, and (a) $\Omega = 0.65$, (b) $\Omega = 0.74$, (c) $\Omega = 0.76$, (d) $\Omega = 0.78$ obtained with the one-vortex initial state (\ref{fun}). Panels (e), (f), (g), and (h) display these vortex lattices, respectively, obtained with the Gaussian initial state. The employed trap parameters are $ \nu=\gamma=1$, the space steps are DX=DY=0.05, the time step is 0.00025 and the space grid is $257\times 257$. 
}
\label{fig4}
\end{figure} 
}

\begin{figure}[!t] 
\centering
\includegraphics[height=.18\linewidth]{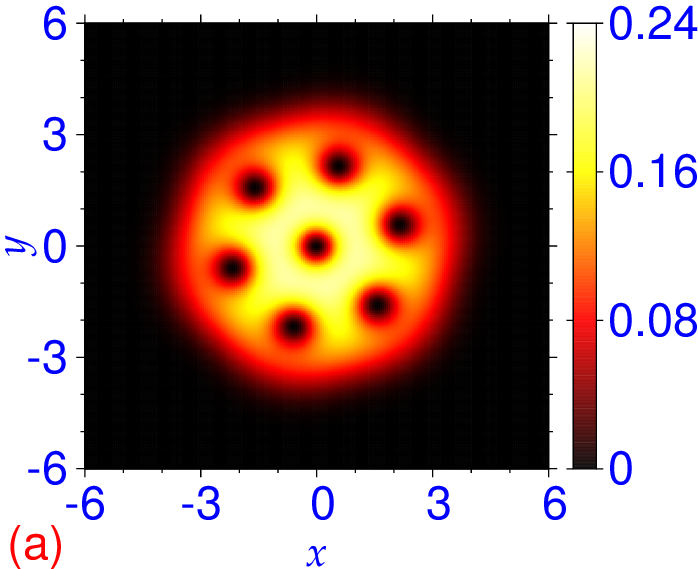}\hspace*{1mm}
\includegraphics[height=.18\linewidth]{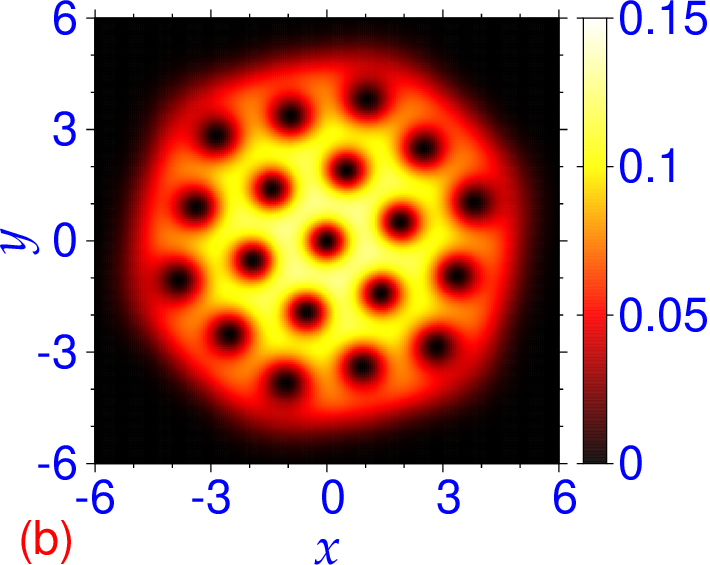}\hspace*{1mm}
\includegraphics[height=.18\linewidth]{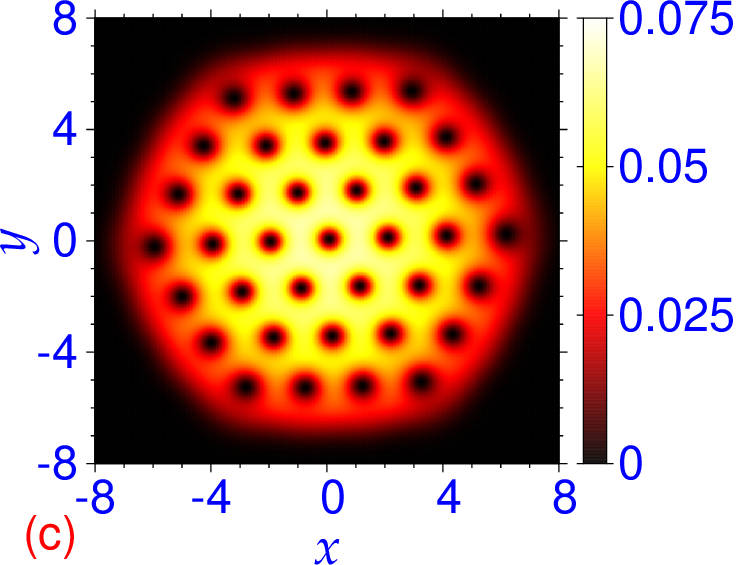}\hspace*{1mm}
\includegraphics[height=.18\linewidth]{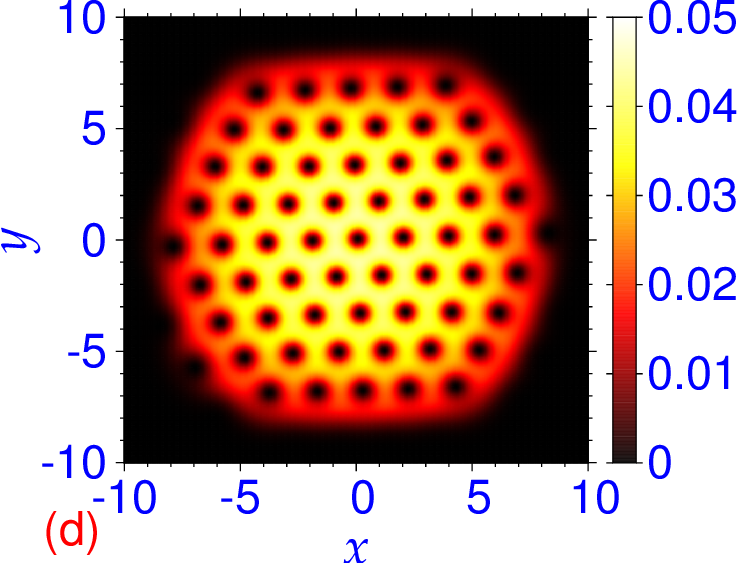}\vspace*{1mm}
\includegraphics[height=.18\linewidth]{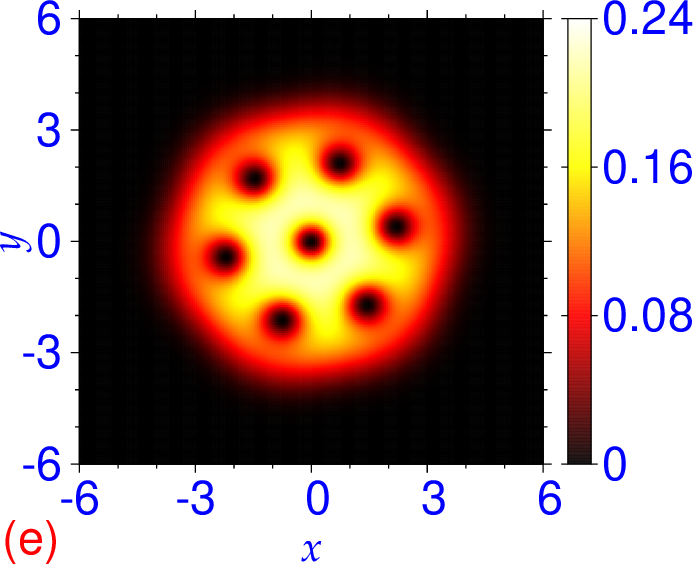}\hspace*{1mm}
\includegraphics[height=.18\linewidth]{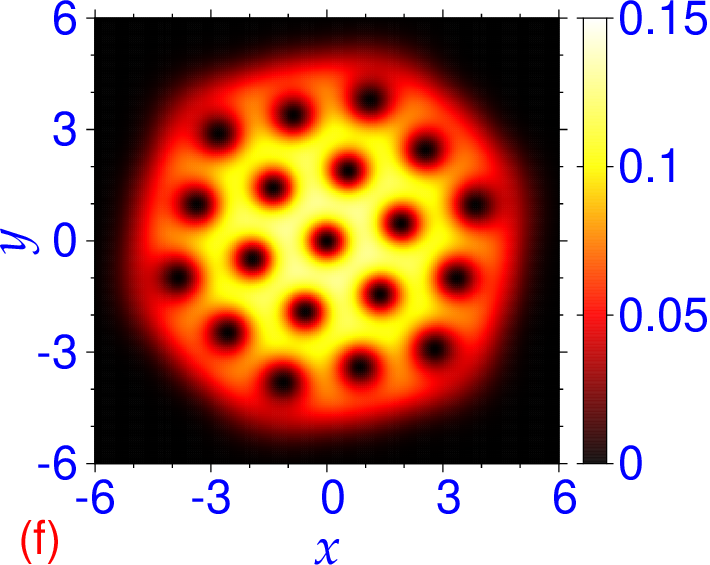}\hspace*{1mm}
\includegraphics[height=.18\linewidth]{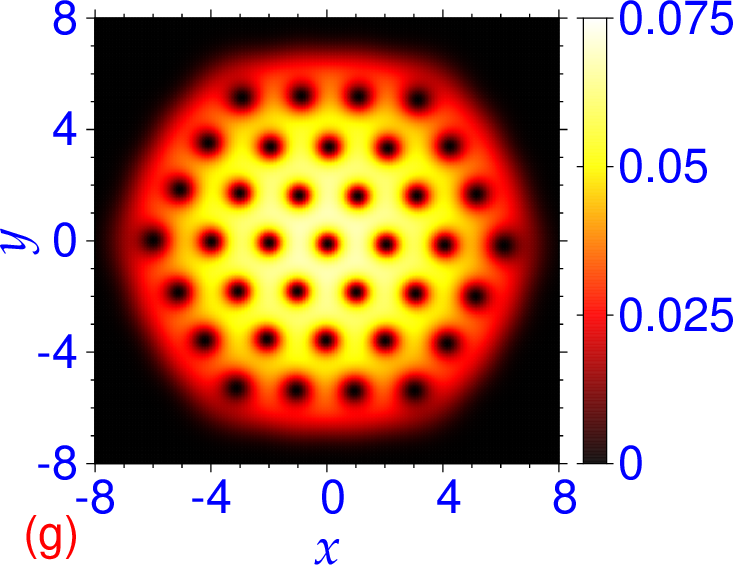}\hspace*{1mm}
\includegraphics[height=.18\linewidth]{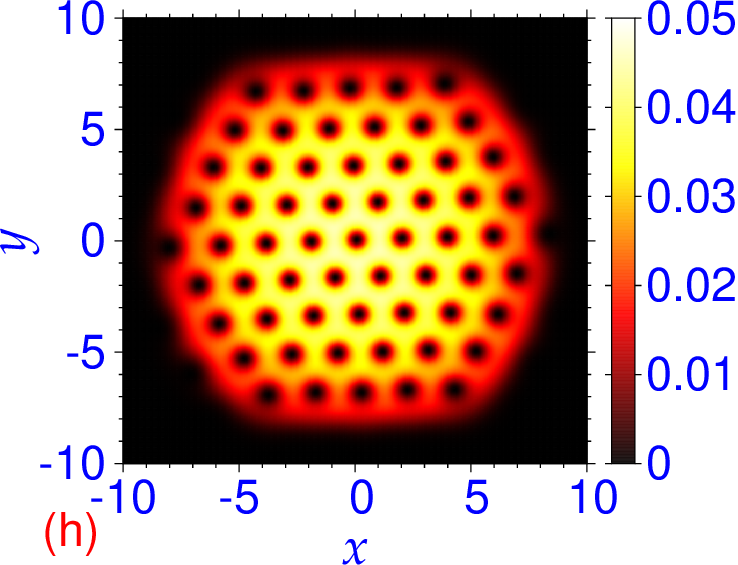} 
\caption{Contour plots of the density profiles for the generated vortex lattices by the 3D imaginary-time propagation of Eq.~(\ref{3d}) for: (a) $ g_{2D}=100$, $ g_{3D}\equiv g_{2D} \sqrt{2\pi/\lambda}= {25.0662827}$, $\Omega = 0.8$, (b) $g_{2D}=100$, $g_{3D} = 25.0662827$, $\Omega = 0.95$, (c) $g_{2D}=500$, $ g_{3D}= 125.33141$, $\Omega=0.92$, (d) $g_{2D}=500$, $ g_{3D}= 125.33141$, $\Omega = 0.978$. Panels (e), (f), (g), and (h) display these vortex lattices, respectively, after the additional real-time propagation for 100 units of time using the corresponding imaginary-time wave function as input. The employed trap parameters are $\nu=\gamma=1$, $\lambda=100$, the space steps are DX=DY=0.05, DZ=0.025, and the time steps are 0.00025 in imaginary time and 0.0001 in real time. The space grids used are: (a) $257\times 257\times 33$, (b) $321\times 321\times 33$, (c) $401\times 401\times 33$, (d) $451\times 451\times 33$.
}
\label{fig5}
\end{figure} 

In Figs.~\ref{fig2}(b)-(d) we illustrate 2D vortex lattices with 19, 37, 
and 61 vortices, respectively, arranged in triangular lattices in the shape of a closed hexagonal form obtained with parameters $g_{2D}=100$, $\Omega = 0.95$ in \ref{fig2}(b), $g_{2D}=500$, $\Omega = 0.92$ in \ref{fig2}(c), and $g_{2D}=500$, $\Omega = 0.978$ in \ref{fig2}(d). As already suggested above, the vortex lattices of Figs.~\ref{fig2}(b), (c), and (d) were obtained using the final wave functions of Fig.~\ref{fig2}(a), (b), and (c), respectively, as the initial states, to speed up the convergence. 
We demonstrate the stability of the obtained vortex lattices using the real-time propagation for 500 time units in Figs.~\ref{fig2}(e)-(h).

{ In Figs.~\ref{fig4} we show the increase of the number of vortices with the increase of the angular frequency $\Omega$ for a
 fixed $g_{2D}=100$ as obtained with the one-vortex initial function and the Gaussian initial function, both modulated by a random phase at different space points. The number of vortices and their orientation in space are identical with both functions, although the energy varies a little from one initial function to another. If the random-phase modulation is removed, these two functions lead to different number of vortices, whereas with the random-phase modulation these functions usually lead to the same number of vortices, viz.~Fig.~\ref{fig4}. 
}

In Figs.~\ref{fig5} we present the $z$-integrated reduced 2D density $\int dz\, |\phi(x,y,z)|^2$,
calculated from the 3D imaginary-time runs, with 7,19, 37, and 61 vortices for the parameters:
(a) $g_{2D}=100$, $\Omega=0.8$, (b) $g_{2D}=100$, $ \Omega= 0.95$, (c) $g_{2D}=500$, 
$\Omega = 0.92$, (d) $g_{2D}=500$, $\Omega = 0.978$. The vortex lattices of Figs.~\ref{fig5}(b)-(d) were generated, as before, by the imaginary-time propagation of Eq.~(\ref{3d}) until the convergence using the final wave function of Figs.~\ref{fig5}(a)-(c) as the initial states, respectively. 
Figures~\ref{fig5}(e)-(h) illustrate the same reduced densities obtained from the 3D real-time runs after 100 time units using as inputs the final converged imaginary-time wave function of Figs.~\ref{fig5}(a)-(d), respectively. The agreement between the imaginary- and the real-time densities demonstrates the stability of the vortex-lattice structures and the employed algorithm. The 2D densities of Fig.~\ref{fig5} are quite similar to those in Fig.~\ref{fig2} with the same 2D nonlinearity and the same angular frequency. To the best of our knowledge, such a clean 61-vortex lattice, viz.~Fig.~\ref{fig5}(d), is obtained for the first time here in the simulation of the 3D GP equation (\ref{3d}).
 
\begin{table}[tp]
\caption{Energy $E$ and chemical potential $\mu$ for the rotating BECs in 2D and 3D shown in Figs.~\ref{fig2} and \ref{fig5}, respectively. For parameters $g_{2D}=100$, $\Omega =0.8$ the calculation is performed with the initial state (\ref{fun}). For the BECs from panels (b), (c), and (d) 
in Figs.~\ref{fig2} and \ref{fig5} the calculation is performed with the converged wave functions
of the corresponding panels (a), (b), and (c) as the initial states.
}
\label{tab1}
\centering
\begin{tabular}{ccccccc}
\hline
 & $g_{2D}=100$ &$g_{2D}=100$ &$g_{2D}=500$ &$g_{2D}=500$ &\\
 & $\Omega=0.8$ &$\Omega=0.95$ & $\Omega=0.92$ & $\Omega=0.978$ & \\
\hline
$\mu$ (2D) &4.351 &2.871 &6.257& 4.198 &\\
$E$ (2D) &3.190 & 2.209& 4.424& 2.951 &\\
 $\mu$ (3D) &54.32 &52.85 & 56.20 &54.17 & \\
 $E$ (3D) & 53.17& 52.19 & 54.40 &52.94 & \\ 
\hline
\end{tabular}
\end{table}

In Table~\ref{tab1} we show the energy and the chemical potential of the BECs of Figs.~\ref{fig2}(a) and \ref{fig5}(a) calculated starting from the analytic function (\ref{fun}) as the initial state. We also give the energy and the chemical potential of the BECs of Figs.~\ref{fig2}(b)-(d) and \ref{fig5}(b)-(d),
calculated with the converged wave functions of Figs.~\ref{fig2}(a)-(c) and \ref{fig5}(a)-(c), respectively, as 
the initial states. The 2D energy values $E= 3.190$ and 2.209 shown in Table~\ref{tab1} for $g_{2D}=100$
and $\Omega =0.8$ and 0.95, respectively, are in good agreement with the energies 
 $E= 3.1904$ and 2.2106 reported in Fig.~6 of Ref. \cite{jeng}. The authors of Ref. \cite{bao}
also calculated the 2D energy and the chemical potential and we verified using the same parameters that the present energies and chemical potentials are in qualitative agreement with their calculations. 

\begin{figure}[!b]
\centering
\includegraphics[height=.25\linewidth]{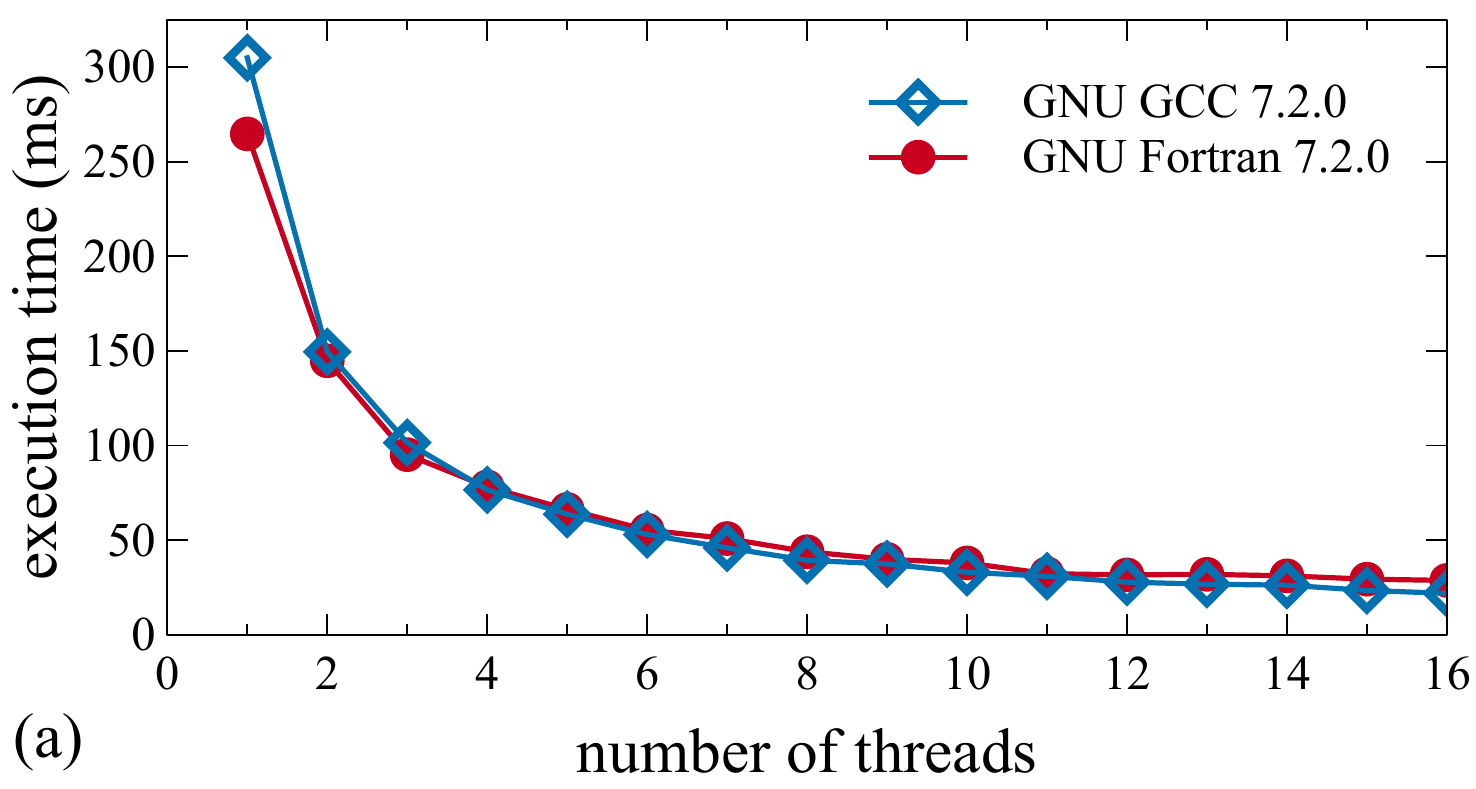}\hspace*{1mm}
\includegraphics[height=.25\linewidth]{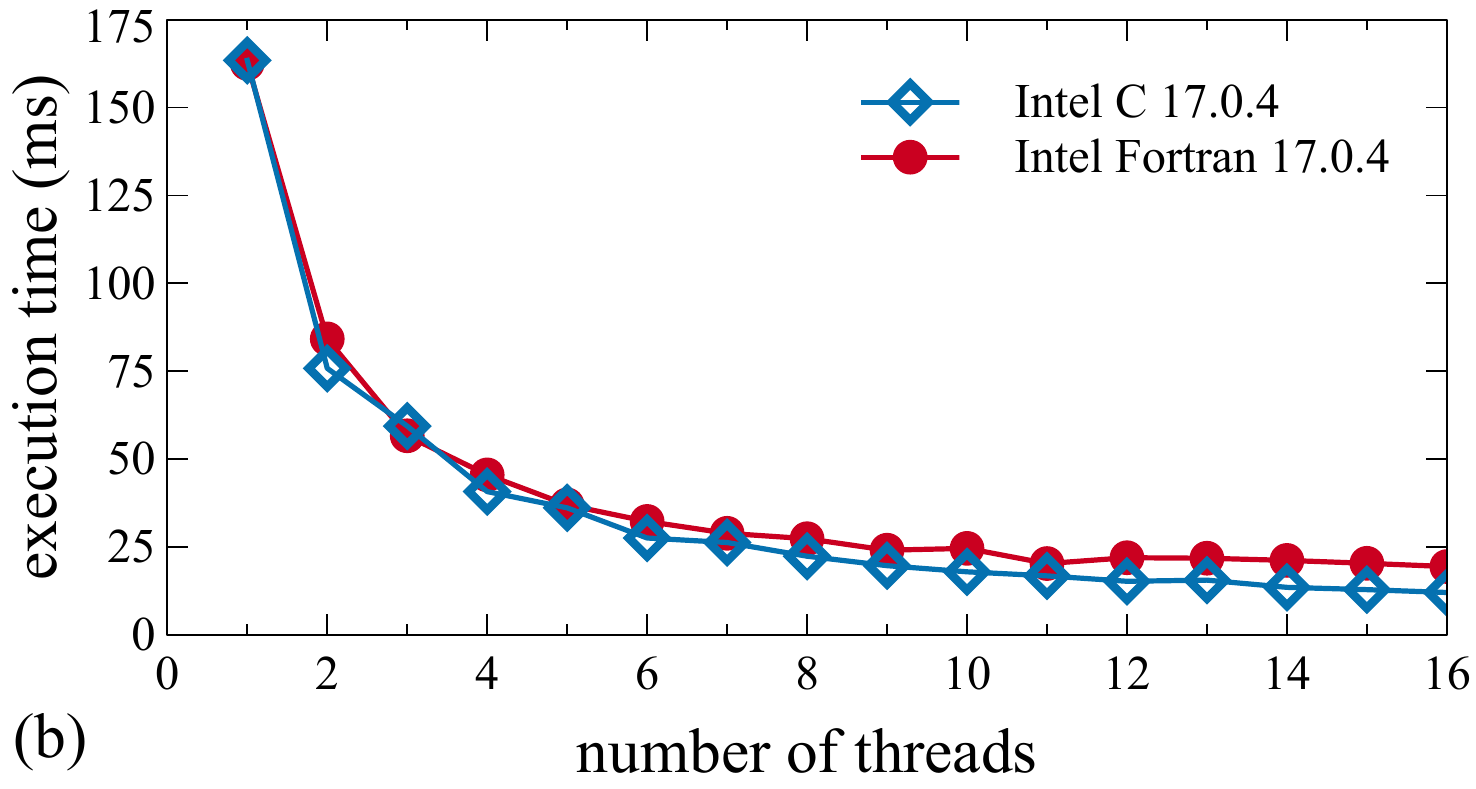}
\caption{Wall-clock execution times of BEC-GP-ROT-OMP programs for imaginary-time propagation in 3D (bec-gp-rot-3d-th), compiled with (a) GNU compiler and (b) Intel compiler, as functions of the number of OpenMP threads. The execution times given here are for one iteration, calculated as averages using runs with 1000 iterations (in milliseconds, excluding initialization and input/output operations, as reported by each program) and with the grid size $257\times 257\times 33$.
}
\label{fig6}
\end{figure} 

We have tested the performance and scalability of our programs on a modern 8-core Intel Xeon E5-2670 CPUs with 32 GB of RAM. The nodes used for testing contain two CPUs, which allowed us to study the performance of our programs on up to 16 CPU cores. The testing was done at the PARADOX supercomputing facility of the Institute of Physics Belgrade.

For both the C and the Fortran programs the execution time in the beginning reduces rapidly as the number of threads (used CPU cores) is increased. But eventually the gain in the execution time saturates. This is illustrated in Fig.~\ref{fig6}, where we plot the execution time versus the number of threads for both the C and the Fortran programs using GNU 7.2.0 and Intel 17.0.4 compilers, respectively. For both compilers, for a large number of threads the C programs are faster. For a small number of threads (four or less), the Fortran programs compiled with the GNU compiler are faster, whereas for the Intel compiler all programs have similar performance, with the C programs being slightly faster.
 
\begin{figure}[!t] 
\centering
\includegraphics[height=.25\linewidth]{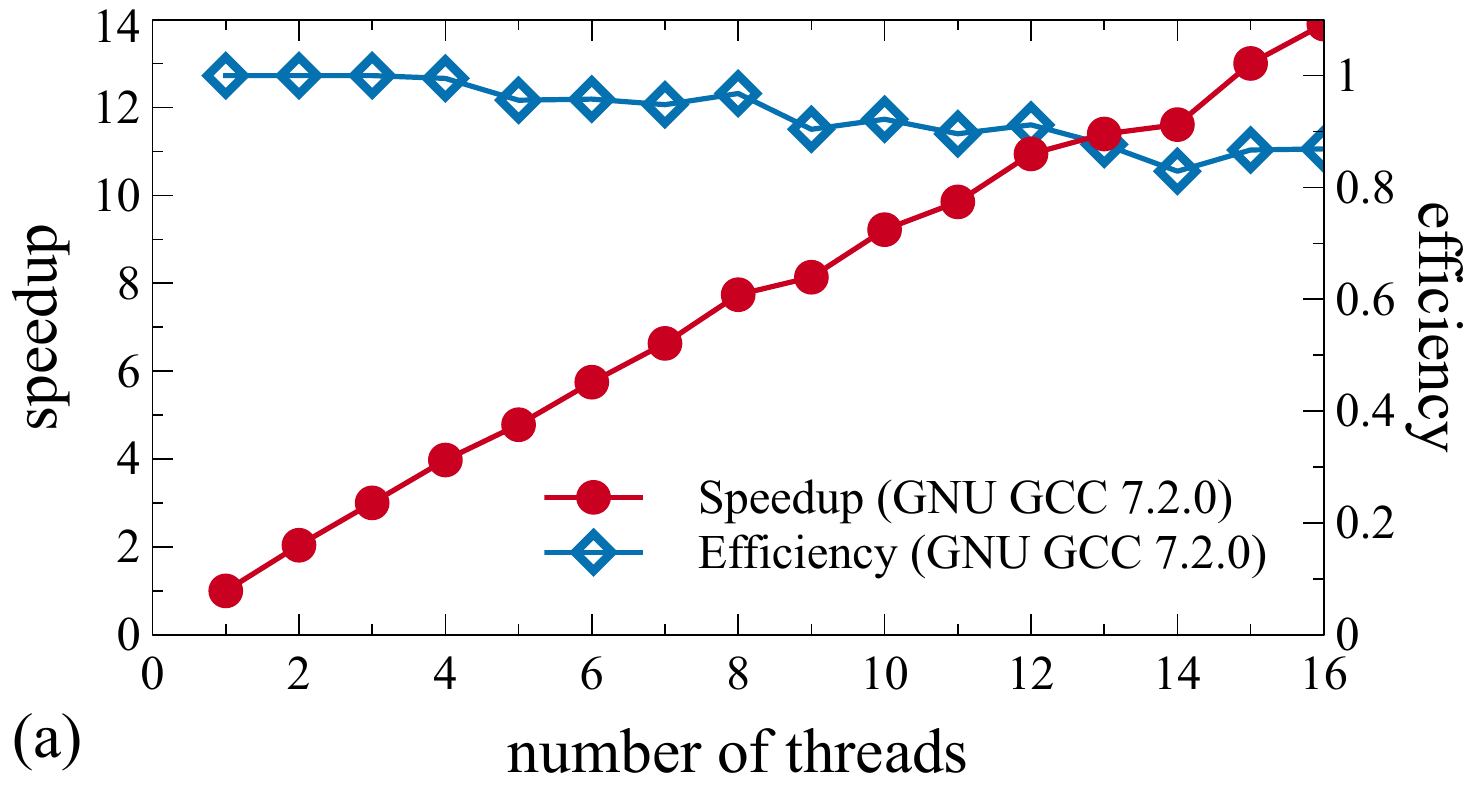}\hspace*{1mm}
\includegraphics[height=.25\linewidth]{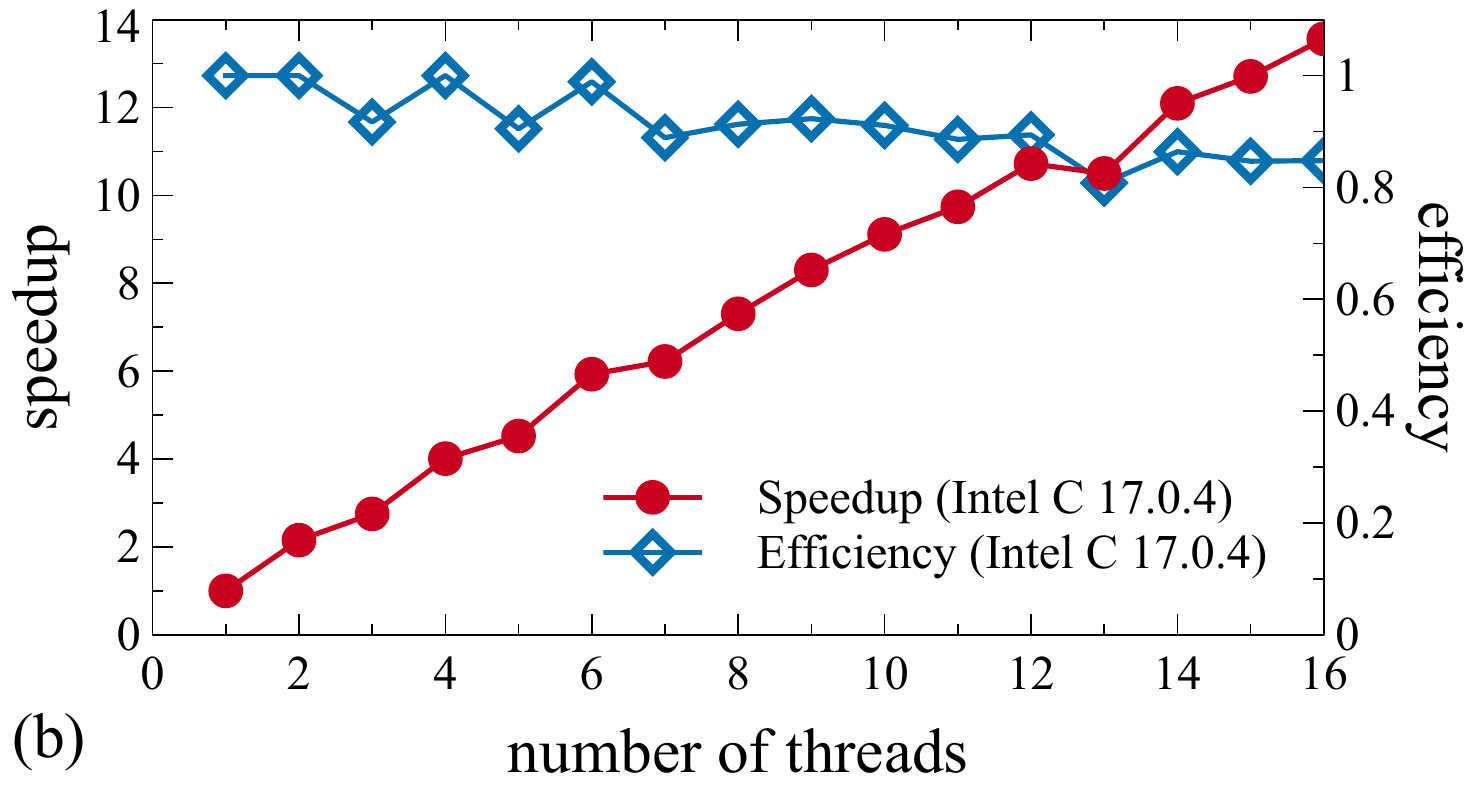}
\includegraphics[height=.25\linewidth]{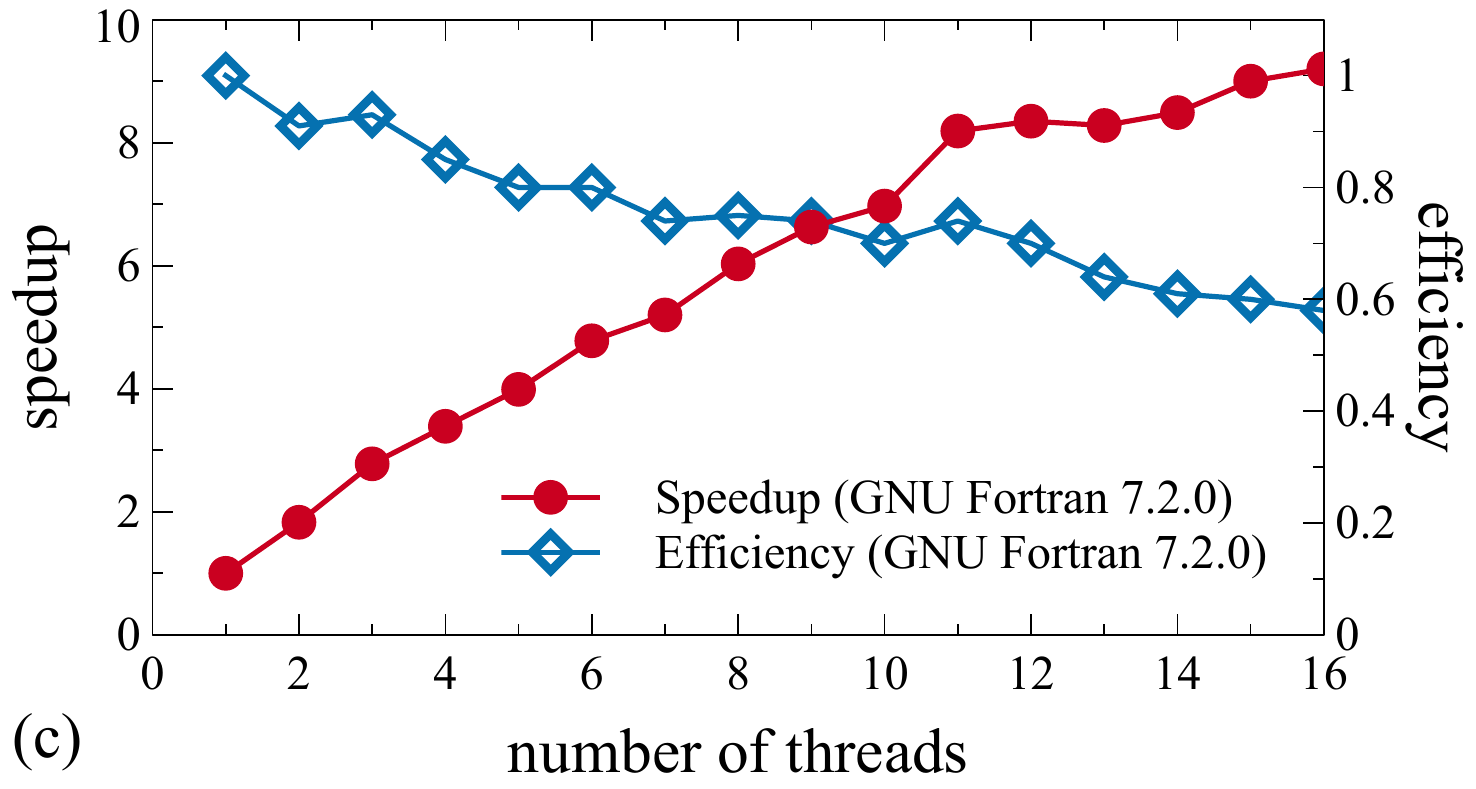}\hspace*{1mm}
\includegraphics[height=.25\linewidth]{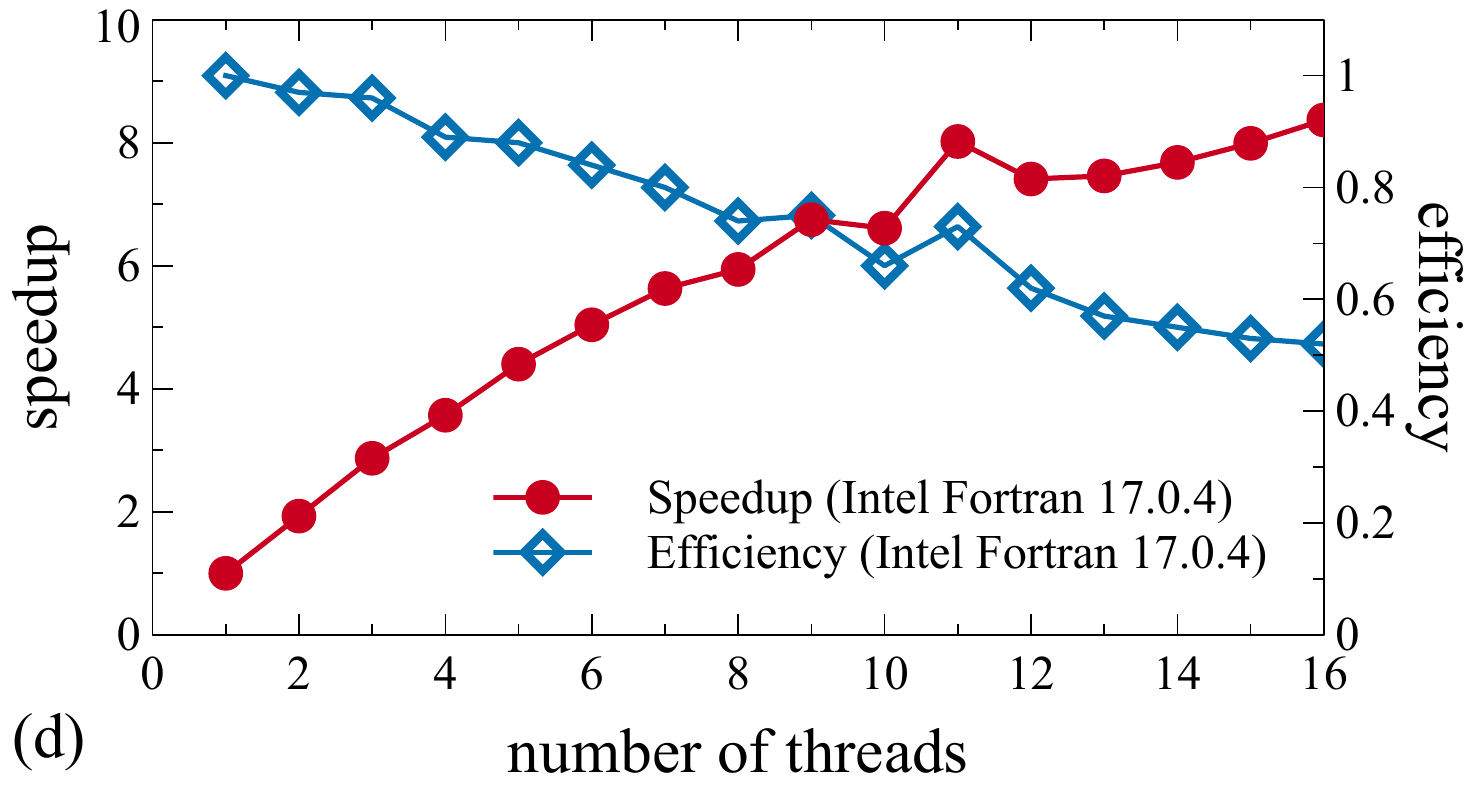}
\caption{Speedup in the execution time and scaling efficiency of BEC-GP-ROT-OMP programs for imaginary-time propagation in 3D (bec-gp-rot-3d-th), compared to single-threaded runs for: (a) C program compiled with the GNU compiler, (b) C program compiled with the Intel compiler, (c) Fortran program compiled with the GNU compiler, (d) Fortran program compiled with the Intel compiler. The speedup is calculated as a ratio of the wall-clock execution times $T(1)/T(n)$ for a single-threaded run and a run with $n$ threads, and the scaling efficiency is calculated as a fraction of the obtained speedup compared to a theoretical maximum $n$. Grid size used for testing is $257\times 257\times 33$.
}
\label{fig7}
\end{figure}

For a quantitative estimate of the performance we now study the speedup and the efficiency of the programs using different compilers for a calculation: GNU GCC 7.2.0, Intel C 17.0.4, GNU Fortran 7.2.0, and Intel Fortran 17.0.4. 
The speedup is defined as the ratio $T(1)/T(n)$ where $T(n)$ is the execution time of a run with $n$ threads. The efficiency is the ratio $T(1)/[nT(n)]$, indicating how many of the threads the computer is effectively utilizing.
These are illustrated in Fig.~\ref{fig7} for GNU GCC, Intel GCC, GNU Fortran, and Intel Fortran compilers, respectively. For a large number of threads, the C programs, viz.~plots \ref{fig7}(a)-(b), are more scalable, with large speedup and efficiency compared to the Fortran programs, viz.~plots \ref{fig7}(c)-(d). The programs in both programming languages are quite efficient and optimized, but a user should use the 
specific program and compiler with which he/she has more experience and feels more comfortable.

\section{Summary and conclusions}
\label{sec:con}

We have presented the efficient OpenMP C and Fortran programs for solving the GP equation for a rotating BEC and use them to calculate the vortex lattices of a rotating BEC by solving the GP equation in the rotating frame. We provide two sets of programs $-$ one for a 3D BEC and the other 
for a quasi-2D BEC. Each of these programs is capable of executing both the imaginary- and the real-time propagation.
We use the split-step Crank-Nicolson algorithm and the programs are based on our earlier OpenMP C and Fortran programs of Ref.~\cite{bec2017x} for a non-rotating BEC. We solve the GP equation by the imaginary-time propagation with the analytic wave function (\ref{fun}) as the initial state to generate a vortex lattice with a small number of vortices. To solve the GP equation with a large number of vortices it is much more efficient to use 
a converged wave function with a smaller number of vortices as the initial state, rather than the analytic function (\ref{fun}). However, the solution can be obtained with any initial state.
Nevertheless, the convergence with one initial state could be much faster than with another initial state. For example, to solve the 2D GP equation (\ref{2d}) with parameters $g_{2D}=100$ and 
$\Omega=0.95$ by the imaginary-time propagation using the initial function (\ref{fun}) and obtain the vortex lattice with 19 vortices, one needs {$12\times 10^5$} time iterations, viz.~Fig.~\ref{fig3}. For the same calculation using the pre-calculated vortex lattice with 7 vortices it is sufficient to use only 30,000 time iterations.
Although both the C and the Fortran programs produce equivalent results, on a multi-core computer with more than 8 cores, the C programs compiled with both the GCC and the Intel compiler yield a more efficient and faster performance.

The localized normalizable initial function (\ref{fun}) has a random phase at each grid point $(x,y)$ which is necessary to obtain a converged vortex lattice with any number of vortices $-$ even or odd $-$ independent of the initial function. If the random phase is removed from the initial function, the one-vortex initial function (\ref{fun}) leads to a vortex lattice with an odd number of vortices and a Gaussian initial function leads to a vortex lattice with an even number of vortices. Any localized normalizable initial function with random phase as in Eq.~(\ref{fun}), e.g., a Gaussian function or a function with one vortex, usually leads to the same vortex lattice. Without the random phase these functions lead to different vortex 
lattices.

\section*{Acknowledgements}
\noindent
V.L. and A.B. acknowledge support by the Ministry of Education, Science, and Technological Development of the Republic of Serbia under projects ON171017 and III43007.
P.M. acknowledges support by the Council of Scientific and Industrial Research (CSIR), Govt. of India under project No. 03(1422)/18/EMR-II. 
R.K.K. acknowledges support by the FAPESP (Brazil) grant 2014/01668-8.
S.K.A. acknowledges support by the CNPq (Brazil) grant 303280/2014-0, by the FAPESP (Brazil) grant 2012/00451-0, and by the ICTP-SAIFR-FAPESP (Brazil) grant 2016/01343-7.
Numerical tests were partially carried out at the PARADOX supercomputing facility at the Scientific Computing Laboratory of the Institute of Physics Belgrade.

\end{document}